\documentclass[traditabstract]{aa}
\usepackage{graphicx,epsf,textcomp,pstricks,psfrag}
\usepackage{txfonts}
\usepackage{longtable,booktabs}
\usepackage{natbib}
\bibpunct{(}{)}{;}{a}{}{,}

\begin{document}

\sloppy


\newcommand{\avk}[1]{{\bf [#1 -- AVK.]}}
\newcommand{\mr} [1]{{\bf [#1 -- MR.]}}
\newcommand{\sm} [1]{{\bf [#1 -- SM.]}}
\newcommand{\rn} [1]{{\bf [#1 -- RN.]}}
\newcommand{\rev}[1]{#1}
\newcommand{\Rev}[1]{#1}

\newcommand{\be}{\begin{equation}}
\newcommand{\ee}{\end{equation}}
\newcommand{\bd}{\begin{displaymath}}
\newcommand{\ed}{\end{displaymath}}
\newcommand{\bea}{\begin{eqnarray}}
\newcommand{\eea}{\end{eqnarray}}
\newcommand{\etal}{et al.}

\newcommand{\mum}{\,\mu\hbox{m}}
\newcommand{\mm}{\,\hbox{mm}}
\newcommand{\cm}{\,\hbox{cm}}
\newcommand{\m}{\,\hbox{m}}
\newcommand{\km}{\,\hbox{km}}
\newcommand{\AU}{\,\hbox{AU}}
\newcommand{\second}{\,\hbox{s}}
\newcommand{\yr}{\,\hbox{yr}}
\newcommand{\g}{\,\hbox{g}}
\newcommand{\kg}{\,\hbox{kg}}
\newcommand{\rad}{\,\hbox{rad}}
\newcommand{\erg}{\,\hbox{erg}}

\title{A possible architecture of the planetary system HR~8799}

\bigskip
\author{Martin Reidemeister\inst{1}
        \and
        Alexander V. Krivov\inst{1}
        \and
        Tobias O. B. Schmidt\inst{1}
        \and
        Simone Fiedler\inst{1}
        \and
        Sebastian M\"uller\inst{1}
        \and\\
        Torsten L\"ohne\inst{1}
        \and
        Ralph Neuh\"auser\inst{1}
       }
\offprints{A.V.~Krivov, \email{krivov@astro.uni-jena.de}}
\institute{Astrophysikalisches Institut, Friedrich-Schiller-Universit\"at Jena,
           Schillerg\"a{\ss}chen~ 2--3, 07745 Jena, Germany
          }
\date{Received {\em March 12, 2009}; accepted {\em May 20, 2009}}

\abstract
{
HR8799 is a nearby A-type star with a debris disk and
three planetary candidates recently imaged directly.
We undertake a coherent analysis of various portions of observational data
on all known components of the system, including the central star, imaged companions, and dust.
The goal is to elucidate the architecture and evolutionary status of the system.
We try to further constrain the age and orientation of the system, orbits and masses of
the companions, as well as the location of dust.
From the high luminosity of debris dust and dynamical constraints,
we argue for a rather young system's age of $\la 50$~Myr.
The system must be seen nearly, but not exactly, pole-on.
Our analysis of the stellar rotational velocity yields
an inclination of $13$ -- $30^{\circ}$, whereas $i \ga 20^\circ$ is needed
for the system to be dynamically stable, which suggests
a probable inclination range of $20$ -- $30^{\circ}$.
The spectral energy distribution, including the Spitzer/IRS spectrum in the
mid-infrared as well as IRAS, ISO, JCMT and IRAM observations, is naturally reproduced
with two dust rings associated with two planetesimal belts.
The inner ``asteroid belt'' is located at $\sim 10$~AU
inside the orbit of the innermost companion and a ``Kuiper belt'' at $\ga 100$~AU
is just exterior to the orbit of the outermost companion.
The dust masses in the inner and outer ring
are estimated to be $\approx 1 \times 10^{-5}$ and $4 \times 10^{-2}$ Earth masses, respectively.
We show that all three planetary candidates may be stable in the mass range
suggested in the discovery paper by Marois et al. 2008
(between 5 and 13 Jupiter masses), but only for some of all possible orientations.
For $(M_b,M_c,M_d) = (5,7,7)$ Jupiter masses, an inclination $i \ga 20^\circ$ is required
and the line of  nodes of the system's symmetry plane on the sky must lie within
$0^\circ$ to $50^\circ$ from north eastward.
For higher  masses $(M_b,M_c,M_d)$ from $(7,10,10)$ to $(11,13,13)$, the constraints on both angles
are even more stringent.
Stable orbits imply a double (4:2:1) mean-motion resonance between
all three companions.
We finally show that in the cases where the companions themselves are orbitally stable,
the dust-producing planetesimal belts are also stable against
planetary perturbations.


}

\authorrunning{Reidemeister et al.}
\titlerunning{The planetary system HR~8799}

\maketitle

\section{Introduction}

HR~8799 is an A5V star located at $\approx$ 40~pc away from Earth, around which
three planetary candidates\footnote{For the sake of brevity,
we often call them ``planets'' throughout this paper.
This comes with the warning that the often used definition
of an ``extrasolar planet'' as a star-orbiting body with mass under
the deuterium burning limit remains controversial. Furthermore, it is not possible
at present to completely exclude the possibility that the mass of at least one companion
in the HR8799 system lies above that limit, although this appears rather unlikely.}
have recently been imaged \citep{marois-et-al-2008}.
All three objects have been shown to be co-moving with the star.
For two of them, also a differential proper motion consistent
with the orbital motion of a companion around the star has been detected.
The presence of the outermost companion has recently
been confirmed by \citet{lafreniere-et-al-2009} through analysis of
archival HST/NICMOS data from 1998
and by \citet{fukagawa-et-al-2009} with SUBARU/CIAO data from 2002.
No further companions with masses greater than $3 M_{\rm Jup}$ (Jupiter masses)
within 600~AU from the star have been found \citep{close-males-2009}.
Three imaged companions are located at projected distances from 24 to 68 AU
from the star and presumably have masses of the order of 7--10 Jupiter masses,
although the mass determination is based solely on untested
evolutionary models and an assumption that the system's age lies in
the range 30--160~Myr.
First dynamical analyses
\citep{fabrycky-murrayclay-2009,gozdziewski-migaszewski-2009} show that,
with the current estimates of planetary masses and the assumption that true orbital separations
of the planets are close to projected distances, the stability of the system on
timescales comparable to the stellar age is not obvious. Still, the system can be
stable, as it likely is, for instance if the planets are locked
in resonances and/or the planetary masses are lower than estimated.

Apart from planets, HR~8799 has long been known to harbor
cold circumstellar dust responsible for excess emission in the far-infrared discovered by IRAS
\citep{sadakane-nishida-1986,zuckerman-song-2004,rhee-et-al-2007}.
The rather strong infrared excess has been also confirmed with ISO/ISOPHOT
measurements \citep{moor-et-al-2006}. Additionally,
Spitzer/IRS measurements revealed warm dust emission in the mid-infrared
\citep{jura-et-al-2004,chen-et-al-2006}. Both cold and warm dust emission must be
indicative of one or more dust-producing planetesimal belts, similar
to the Edgeworth-Kuiper belt and possibly, the asteroid belt in the solar system.
Altogether, there appears to be an emerging view of a complex, multi-component
planetary system with several planets, planetesimal belts, and dust.

Given the fact that the planets have been discovered very recently,
it is not surprising that our knowledge of the system is very poor and that
even the key parameters of the system and its components remain vaguely known.
A large uncertainty in the system's age amplifies the difficulty of
inferring accurate masses of the companions from evolutionary
models, and the mass estimates vary from one model to another
even for the same age.
While there are clear indications that the system is
seen nearly pole-on, the exact orientation of its symmetry plane
is not known either, which makes it impossible to convert projected
astrocentric distances of planets into their true positions.
The differential proper motion was measured with a reasonable accuracy
only for the outer planets, and even in that case the accuracy is not
yet sufficient to constrain orbital eccentricities. As far as the dust
is concerned, the debris disk remains unresolved, offering no possibility to
better recover the orientation of the system's plane. Even the photometry
data are scarce.
This results in a poor knowledge of the dust distribution.

Obviously, most of these issues could only be resolved with new observational data,
some of which may come promptly, while some others may require longer time
spans. However, already at the current stage, we find it reasonable to
reanalyze the available data. While the discovery paper by
\citet{marois-et-al-2008} mostly concentrates on the planets themselves
and the first ``follow-up'' publications
\citep{fabrycky-murrayclay-2009,gozdziewski-migaszewski-2009}
provide an in-depth analysis of dynamical stability issues,
in this paper we try to paint a more synthetic
view of the planetary system around HR~8799 with all its components~---
central star, planets, and dust-producing planetesimal belts.

Section 2 focuses on the stellar properties, notably the stellar age
(section 2.1)  and inclination of the rotation axis, assumed to
coincide with the inclination of the whole system (section 2.2).
In Section 3, we analyze the observed spectral energy distribution (SED, section 3.1) and try to
fit the data with several dust belts (section 3.2).
Section 4 treats presumed planets and tries to constrain their masses
both from evolutionary models (section 4.1) and dynamical stability requirement
(section 4.2).
Section 5 checks whether planetesimal belts, as found to fit the infrared photometry,
would be dynamically stable against planetary perturbations.
Section 6 contains our conclusions and a discussion.

\section{The central star}

\subsection{Age}

\citet{marois-et-al-2008} give an age of 30 to 160 Myr for HR 8799 and hence
its companions, considering several age indicators.
While most of the indicators
support a rather younger age, few of them still allow for an older age or even
suggest it. One indicator is the galactic space motion (UVW) of the primary, which
is similar to that of close young associations \citep{marois-et-al-2008}. Using these
data, \citet{moor-et-al-2006} propose HR 8799 to be a member of the Local Association at
an age of 20 to 150 Myr with a probability of 62\,\%. Another method is the position
of HR 8799 in the Hertzsprung-Russell diagram.
Taking into account the low luminosity of the star (after correction for
its low metallicity) as well as the UVW space motion,
\citet{rhee-et-al-2007} arrived at an age of 30~Myr.

\citet{marois-et-al-2008} further note that $\lambda$ Bootis
stars are generally thought to be young, up to a few 100 Myr.
However, the Hipparcos mission has shown that the well established
$\lambda$ Bootis stars of the Galactic field
comprise the whole range from the Zero Age Main Sequence to the Terminal Age Main Sequence,
which is $\sim$\,1\,Gyr for an A-type star
\citep[][and references therein]{turcotte-2002,paunzen-et-al-2001}.
The best indicator for
a rather older age is the location of HR 8799 in a $T_\mathrm{eff}$ vs. $\log g$ diagram derived
from published uvby$\beta$ photometry. Using this method,
\citet{song-et-al-2001} find an age of 50 to 1128~Myr with a best estimate of 732 Myr
and \citet{chen-et-al-2006} an age of 590 Myr.

An independent argument in favor of a rather younger age may come from the
the dust portion of the system.
The measured infrared excess ratio of $\sim 100$ at $60$~--$90\mum$
(see Fig.~\ref{fig_with photosphere_7400_smooth.eps} below)
would be typical of a debris disk star with the age of $\la 50$~Myr
\citep[see][their Fig. 5]{su-et-al-2006}. However, this argument is purely
statistical and must be taken with caution. For instance, one cannot
exclude the possibility that the formation of such a planetary system with
three massive planets in very wide orbits could trace back
to an exceptionally dense and large protoplanetary disk. The latter might
leave, as a by-product, a more massive debris disk on the periphery, whose
fractional luminosity might be well above the statistically expected level.

Altogether, there seem to be more arguments to advocate a younger age of the system
of the order of several tens of Myr. On any account,
as pointed out in the discovery paper by \citet{marois-et-al-2008} and
is further discussed in section~4 below, extremely old ages would inevitably
imply high object masses in the brown dwarf range~--- for all
the evolutionary models used to infer the masses.
\citet{fabrycky-murrayclay-2009} have shown that
dynamical stability of such a system is problematic.
It can be stable for masses up to 20 Jupiter masses,
but only for very special orbital configurations.

\subsection{Rotational period and inclination}
\label{subsec: Rotational period and inclination}
As summarized by \citet{sadakane-2006} the Vega-like, $\gamma$ Doradus type
pulsator HR 8799 is showing $\lambda$ Bootis-like abundances. He concludes
that for the case of HR 8799, which is known to be a single star and
associated with a dusty disk, the scenario invoking the process of selective
accretion of circumstellar or interstellar material depleted in refractory
elements, seems to be the favorable explanation for the unusually low
abundances.

The fact that HR 8799 is a $\gamma$ Doradus type pulsator makes it hard to
find an indisputable rotational period. However, several authors give a
value of $\sim$\,0.51\,days \citep[e.g.][]{rodriguez-zerbi-1995}.
\citet{zerbi-et-al-1999} found from a multisite campaign three independent
frequencies (0.5053 d, 0.5791 d, 0.6061 d) and a coupling term between them
(4.0339 d). All of these frequencies could be independent pulsational modes.
However, if one of these frequencies corresponds to the rotational period of
the star, we are able to calculate the inclination $i$ of its rotational axis.

From the possible rotational frequencies (0.5053 d, 0.5791 d, 0.6061 d)
and from the radius of HR 8799 of 1.32\,$R_{\odot}$ \citep{allende-et-al-1999} to
1.6\,$R_{\odot}$ \citep{pasinetti-et-al-2001}, we determine
the possible range of the true rotational velocity $v$ of the star of
$110$--$160\km\second^{-1}$.
These values  agree quite well with the median $v\sin{i}$ of A4--A6 main sequence stars of
$159 \pm 7.2\km\second^{-1}$ \citep{royer-et-al-2007}.
On the other hand, the projected rotational velocity $v\sin{i}$ of HR 8799
was measured by several authors to be between $35.5\km\second^{-1}$ and
$55\km\second^{-1}$ \citep[e.g.][]{kaye-et-al-1998,uesugi-et-al-1982}.
From $v$ and $v \sin i$, we finally derive a possible range of the inclination
of the star of $13^\circ$--$30^\circ$. Note that in the above estimates we
excluded the 4.0339-day period as this would lead to $\sin{i} > 1$.

It can be expected that the rotational equator of the star
and the planetary orbits are all aligned with each other.
Spin-orbit alignment is a common assumption, consistent for instance
with the data of most transiting planets \citep[e.g.][]{fabrycky-winn-2009}.
It has also been confirmed for Fomalhaut and its disk \citep{lebouquin-et-al-2009}.
Nonetheless, a misalignment on the order of several degrees is likely.
It is exemplified by our own system, in which the Jupiter orbit is tilted
by $\sim 7^\circ$ to the solar equator and the orbital planes of giant planets
differ from each other by a few degrees.
With this caveat, we make an assumption of the perfect alignment
throughout the paper. This assumption is particularly needed to reduce
the complexity of our dynamical stability studies presented below.

An additional argument in favor of a low inclination of the
whole system comes from the fact that the measured proper motion of the
companions relative to the star
is also consistent with a nearly pole-on view onto the system
\citep{marois-et-al-2008}. The most recent astrometric analysis by
\citet{lafreniere-et-al-2009} further supports this conclusion.
Their best fits with circular orbits suggest the semimajor axis
$a \sim 68$--$74\AU$ and the inclination $i\sim 13^\circ$--$23^\circ$
for the orbit of the outermost companion, HR~8799~b.

\section{Dust and planetesimal belts}

\subsection{Observed SED}

Table \ref{tab:photometry} lists the catalogs and references used to provide
the optical, infrared, sub-mm and millimeter photometry.
We employed the Hipparcos and Tycho
databases as well as USNO and GSC catalogs to compile the optical
photometry, whereas the near-infrared data were taken from the 2MASS survey.
The mid- and far-infrared photometry is provided by the IRAS and ISO
satellites,
while sub-mm and millimeter data were obtained at JCMT and IRAM.
For transforming the $B$, $V$, $R$, $I$ magnitudes into units of
flux [Jy] we used the standard calibration system of Johnson, whereas in
case of the 2MASS $J$, $H$, $K_s$ bands the calibrations of \citet{cohen-et-al-2003}
were applied.
For the IRAS fluxes a color-correction factor
\citep{beichman-et-al-1988} assuming a black body spectral energy distribution
for a temperature of 5000 K was
employed\footnote{Correction factors are only available over a coarse temperature grid
\ldots, 5000~K, 10000~K.}.

The optical and near-infrared photometry was used to derive a best-fit
photospheric model. We estimated the stellar photospheric fluxes by minimum
$\chi^2$ fitting of NextGen model atmospheres \citep{hauschildt-et-al-1999},
using only bandpasses with wavelengths shorter than $3 \mum$, where no excess emission
is expected. In our search for the best fit we employed a system of
NextGen models with an effective temperature step size of 200 K, with a log $g$ of
4.5, and with solar metallicity\footnote{
Although the metallicity of HR8799 is $-0.47$ \citep{gray-kaye-1999},
this parameter has no important effect on the fitting result.}.
Varying the temperature as well as the
stellar radius, which affects the solid angle dilution factor, we derive a
best-fit temperature of 7400 K and a best-fit radius of 1.34 $R_\odot$,
both in a very good agreement with the results
of \citet{gray-kaye-1999}.

In order to obtain more information on the dust component of the system, we
extracted publicly available IRS data for HR8799 from the Spitzer archive
with the Leopard software ({\sf http://ssc.spitzer.caltech.edu/irs/}).
Those data were taken in December 2003 under AOR 3565568 and originally
published by \citet{chen-et-al-2006}.
Using the post-basic correction data provided by the standard IRS pipeline,
we first subtracted the zodiacal light background estimated by the SPOT software.
Then we joined the datasets for the individual IRS modules and orders and
subtracted a simple Rayleigh-Jeans stellar photosphere in the wavelength
region of interest, assuming now that there is no excess at 8~$\mu$m
where the IRS spectrum starts.
For plotting and fitting purposes, the scatter in the resulting dust spectrum was reduced by taking
averages over 4 neighboring data points.
This increases the signal-to-noise ratio without
lowering the spectral resolution, which remains
constrained by the instrumental value $R\sim 60$--$120$.
Spurious features at the edges of the spectral orders around 14 and 20~$\mum$ were removed.

The resulting spectrum shows a somewhat (by $\approx 40$\%) weaker excess than that by
\citet{chen-et-al-2006}, especially between 20 and 30~$\mu$m.
One possible origin of this discrepancy is the approach used for background subtraction.
In addition, both spectra fall well below the IRAS upper limits at 25~$\mu$m but also by
several $\sigma$ below the IRAS 12~$\mu$m faint-source and point-source measurements.
The easiest way to reduce this scaling uncertainty would be using a good photometry
point in the mid-infrared.
Unfortunately, a 24~$\mu$m Spitzer/MIPS point, which would ideally serve for this
purpose, is not yet available in the Spitzer archive,
although these data have recently been taken (Kate Su, pers. comm).
Whether the maximum at around 11~$\mu$m is a silicate feature is unclear.
Definite conclusions would require
a longer exposure time and a higher spectral resolution.

\begin{figure}
  \begin{center}
  \includegraphics[width=0.48\textwidth]{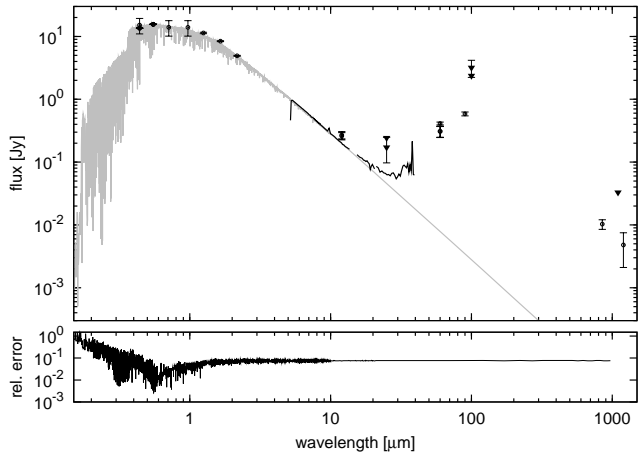}
  \end{center}
  \caption{Spectral energy distribution of HR 8799.
  Circles are photometric measurements, triangles are upper limits
  (Table~\ref{tab:photometry}).
  The black line between $\approx 5$ and $40\mum$ is the Spitzer/IRS spectrum.
  Overlaid is the best-fit NextGen model
  (grey line) for the stellar photosphere.
  Its estimated relative error is plotted under the main panel.
  }
  \label{fig_with photosphere_7400_smooth.eps}
\end{figure}

The resulting set of photometry points and the IRS spectrum are shown in
Fig.~\ref{fig_with photosphere_7400_smooth.eps} (with photosphere) and
in  Fig. \ref{fig_SED_fit} (excess emission only).

\begin{table}[!h]
\caption{Photometry of HR8799. IRAS fluxes are color corrected as
described in the IRAS Explanatory Supplement \citep{beichman-et-al-1988}.}
\label{tab:photometry}

  \centering
  \tiny
  \begin{tabular}{lllr}
  \hline
\hline
Photometric band  & Flux or magnitude     &
$F_{qual}$ (IRAS) & Ref.  \\
\hline
\hline
  & [mag]     &   &   \\
\hline
$B$ & 6.090 $\pm$ 0.300 &   & (1) \\
$B$ & 6.196     &   & (2) \\
$B$ & 6.210 $\pm$ 0.010 &   & (3) \\
$B$ & 6.214 $\pm$ 0.009 &   & (4) \\
$V$ & 5.960 $\pm$ 0.010 &   & (4) \\
$V$ & 5.959     &   & (2) \\
$V$ & 5.960 $\pm$ 0.010 &   & (3) \\
$R$ & 5.810 $\pm$ 0.300 &   & (1) \\
$I$ & 5.690 $\pm$ 0.300 &   & (1) \\
$J$ & 5.383 $\pm$ 0.027 &   & (5) \\
$H$ & 5.280 $\pm$ 0.018 &   & (5) \\
$K_s$ & 5.240 $\pm$ 0.018 &   & (5) \\
\hline
  & [Jy]      &   &     \\
\hline
IRAS PSC 12 $ \mu m$  & 0.267 $\pm$ 0.034 & 3 & (6) \\
IRAS PSC 25 $ \mu m$  & 0.246 $\pm$ 0.000 & 1 & (6) \\
IRAS PSC 60 $ \mu m$  & 0.307 $\pm$ 0.061 & 2 & (6) \\
IRAS PSC 100 $ \mu m$ & 2.376 $\pm$ 0.000 & 1 & (6) \\
IRAS FSC 12 $ \mu m$  & 0.278 $\pm$ 0.036 & 3 & (7) \\
IRAS FSC 25 $ \mu m$  & 0.174 $\pm$ 0.075 & 1 & (7) \\
IRAS FSC 60 $ \mu m$  & 0.311 $\pm$ 0.062 & 3 & (7) \\
IRAS FSC 100 $ \mu m$ & 3.202 $\pm$ 0.977 & 1 & (7) \\
ISO 60 $ \mu m$         & 0.412 $\pm$ 0.021 &   & (8) \\
ISO 90 $ \mu m$         & 0.585 $\pm$ 0.041 &   & (8) \\
\hline
  & [mJy]     &   &           \\
\hline
JCMT 850 $ \mu m$ & 10.3  $\pm$ 1.8   &   & (9) \\
JCMT 1100 $ \mu m$  & $<33$     &     & (10)  \\
IRAM 1200 $ \mu m$  & 4.8 $\pm$ 2.7 &     & (10)  \\
\hline
\hline

\end{tabular}
\begin{flushleft}
References:
(1) The USNO-B1.0 Catalog \citep{monet-et-al-2003};
(2) NOMAD Catalog \citep{zacharias-et-al-2004}, from Tycho-2 Catalog
\citep{hog-et-al-2000};
(3) The Guide Star Catalog, Version 2.3.2
    \citep{lasker-et-al-2008};
(4) The Hipparcos and Tycho Catalogues \citep{perryman-et-al-1997};
(5) 2MASS All-Sky Catalog of Point Sources \citep{skrutskie-et-al-2006};
(6) IRAS catalogue of Point Sources, Version 2.0
\citep{helou-walker-1988};
(7) IRAS Faint Source Catalog, $|b|$ $>$ 10, Version 2.0
\citep{moshir-et-al-1990};
(8) \citep{moor-et-al-2006}
(9) \citep{williams-andrews-2006}
(10) \citep{sylvester-et-al-1996}
\end{flushleft}
\end{table}

\subsection{Interpretation}

To get a rough idea of the location of the dust belt(s) and the amount of
dust in the HR8799 system, we modeled the SED assuming a double power-law
surface number density of dust $N \propto s^{-q} r^{-\xi}$, where $s$ and
$r$ are the dust grain size and distance from the star, respectively.
Keeping in mind that SED interpretation is a degenerate problem and to
decrease the number of free parameters, we restrict ourselves to the case
of $q = 3.5$ and $\xi = 1$. The dust composition was assumed to be
astronomical silicate \citep{Laor-Draine-1993}. As the minimum grain size,
we chose the radiation pressure blowout radius, which amounts to
$\approx 5~\mu$m for astrosilicate. The maximum grain radius, which has
little effect on the results, was arbitrarily set to $1000~\mu$m.

\begin{figure}
  \begin{center}

  \includegraphics[width=0.45\textwidth]{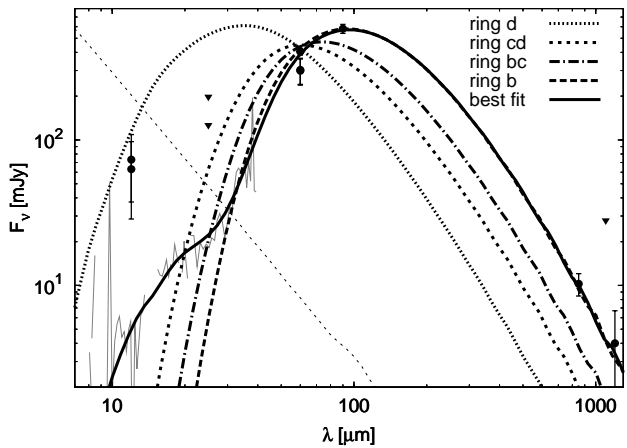}\\
  \end{center}
  \caption{Excess emission of HR 8799 in the infrared.
           The grey solid line is the IRS spectrum.
           The circles are the observed fluxes, and the triangles are upper limits.
           Dotted,  short-dashed, dash-dotted, and long-dashed lines are the SEDs
           from the four first-guess dust rings
           (Table~\ref{tab_ring_first_guess}).
           The black solid line is our `best fit'.}
  \label{fig_SED_fit}
\end{figure}

At first, we modeled the SEDs that would arise from four hypothetical dust
rings  with arbitrarily chosen extensions, located inside the orbit of the
innermost planet HR8799~d, between d and c, between c and b, and outside
the orbit of the outermost planet HR8799~b  (Table \ref{tab_ring_first_guess}).
Dust masses were chosen in such a way as to reproduce the $60~\mu$m flux. The
results are shown in Fig. \ref{fig_SED_fit}.

\begin{table}
  \centering
  \renewcommand{\arraystretch}{1.1}
  \caption{Locations and names of the first-guess dust rings.}
  \label{tab_ring_first_guess}
  \begin{tabular}{lcc}
    \hline\hline
    ring location      & ring extension [AU] & name\\
    \hline\hline
    inside d           &  3 --  15           & ring d \\
    between d and c    & 28 --  32           & ring cd \\
    between c and b    & 45 --  60           & ring bc \\
    outside b          & 75 -- 125           & ring b \\
    \hline\hline
  \end{tabular}
\end{table}

Comparison of the first-guess model SEDs with the available photometry and
spectrometry observations reveals two issues. First, one single ring is
not capable of reproducing the entire set of observations from the mid- to
the far-infrared. Second,
if the $10\mum$ silicate feature in the IRS spectrum is real,
a substantial fraction of particles smaller than the blowout size will be required.

Taking these discrepancies into account, we combined two rings,
`ring d' and `ring b', and fitted the lower size cutoffs and dust masses
to the observations. In our `best fit' model, which is shown as solid line
in Fig. \ref{fig_SED_fit}, the minimum sizes are 2 and $6~\mu$m and the
dust masses are $1.4\times 10^{-5}~M_\oplus$ and
$4.2\times 10^{-2}~M_\oplus$ for the inner and outer ring, respectively.
As a word of caution, we remind that these estimates directly inherit the
uncertainty of the IRS spectrum calibration, as discussed in section 3.1,
which is a factor of several in dust mass.

A question arises whether these fits are physically reasonable. In particular,
it is not immediately clear whether the blowout particles are really needed to reproduce the
observations. Similar problems have already been encountered in studies
of other debris disks \citep[e.g. around Vega, see][]{su-et-al-2005}.
However, such inconsistencies can be overcome by considering a complete
dynamical treatment of the specific disk without significant changes in
the disk location (M{\"u}ller et al., in prep.).
Further, our choice of power-law spatial and size distributions
is only a rough, albeit commonly used, approximation.
More elaborate dynamical studies
\citep[e.g.][]{krivov-et-al-2006,thebault-augereau-2007}
show clear deviations from this assumption, especially in the case of the
dust size distribution where a wavy pattern arises from an underabundance
of small particles induced by radiation blowout.
Thus, we checked the impact of moderate variations in the slopes
($0.1 \leq \xi \leq 1.9$ and $2.5 \leq  q \leq 4.5$). It turned
out that the slope of the spatial distribution $\xi$ has little effect on the results,
moderately changing only the short-wavelength part in the SED from the inner dust ring.
The size distribution slope $q$, however, affects the resulting emission appreciably.
While for the inner ring the changes are still small
(a steeper slope would amplify the silicate features at 10 and $20\mum$
whereas a shallower distribution would wash them out, just as expected),
similar changes in the outer ring would require a strong
compensation by altering other disk parameters (which were fixed in our approach).
The reason is that for the outer component both the rise and the fall of
the SED are well constrained by photometric observations.
They place tight constraints on the width of the SED,
in contrast to the inner ring where only the short-wavelength part
of the SED is known.
Since dust in the outer
ring is much colder than in the inner ring it is no longer the strength
of the features but the width of the SED that is affected by a different
slope in the size distribution: the steeper the distribution,
the narrower the SED.
On any account, for our purpose of
determining the rough location and mass of dust the simple fitting approach used
here is sufficient.

Another question is, how well the edges of the outer and inner ring are
constrained. To check this, we
varied them and fitted the SED again, leaving the lower cutoff
size and dust mass as free parameters. For the outer ring,
we found reasonable fits with the inner edge between 75 and 120~AU
and the outer edge between 125 and 170~AU
(for $q = 3.5$ and $\xi = 1$).
The outer edge in the inner ring can go from 15 down to 10~AU.
In fact, the inner ring truncated at 10~AU provides even slightly better
agreement with the IRS spectrum between 20 and $30~\mu$m. However, due to
the calibration uncertainties it is difficult to assess the accuracy of the
fit, which leaves the outer edge of the inner ring rather unconstrained.
The inner edge of the inner ring can be as close to the star as 2~AU to
conform to the IRS spectrum.

So far we have discussed {\em dust} rings in the system.
However, the presence of a dust belt requires a belt of
{\em planetesimals} that do not show up in the observations, but produce and
sustain visible dust.
Due to the radiation pressure effect, dust grains typically move
in orbits with periastra at the planetesimal belt and apastra outside it.
The smaller the grains, the farther out from the star they are spread.
Thus the dust-producing planetesimal belt is expected to be narrower than
the observed dust ring and to lie within the dust ring close to its inner edge
\citep[e.g.][]{krivov-et-al-2006}.
Therefore, we can expect that the outer planetesimal belt is located
at $\sim 75$--$120$~AU and the inner planetesimal belt at $2$--$10$~AU from
the star.
It is important to check whether the expected locations of the
outer and inner planetesimal belt
are dynamically compatible with the presence
of the outermost and innermost planet, respectively.
We will investigate this in section 5, trying to find additional constraints on
the location of the two rings.

\section{Planets}

\subsection{Masses from models}

A common method to estimate masses of various astrophysical objects,
from stars to planets, is to use their formation and evolution models.
Such models predict
essential physical parameters of objects of various masses,
notably their luminosity and temperature, as a function of age.
If the age is known, a comparison with luminosity or temperature
retrieved from observations allows one to estimate the mass.
In the case of HR8799, the temperature of the companions is unknown,
but their luminosity has been derived from brightness and distance
with sufficient accuracy to apply the models.
Ages, masses, and other parameters of other directly
imaged planet candidates were recently reviewed
by \citet{schmidt-et-al-2009}.

However, all models involve simplifying assumptions
and adopt certain initial conditions, e.g. the initial internal
energy and temperature structure.
Further, the so-called hot-start models all start at a non-zero age
(e.g., 0.1 or 1 Myr)
with a finite luminosity and thus do not cover the actual formation stage.
For these reasons,
models may not deliver reliable results for at least the first several Myrs
\citep[see][for discussion]{wuchterl-2001,chabrier-et-al-2005},
if not several hundred Myrs \citep{stevenson-1982}.

\begin{table*}[hbt!]
\caption{Masses of HR 8799 b / c / d from luminosity (and absolute K-band magnitude) using
various evolutionary models.}
\label{tab_masses_from_models}
\centering
\begin{tabular}{l|ccccccccc}
\hline\hline
HR8799 b & \multicolumn{9}{c}{Luminosity $\log{L/L_{\odot}}= -5.1\,\pm\,0.1$ \citep{marois-et-al-2008}} \\
\hline
Model &  \multicolumn{9}{c}{Mass [M$_{\mathrm{Jup}}$] at age}\\
& 20 Myr & 30 Myr & 60 Myr & 100 Myr & 160 Myr & 590 Myr & 730 Myr & 1000 Myr &  1128 Myr  \\
\hline
 \citet{burrows-et-al-1997} & 3.5 -- 4.5 & 4.5 -- 6 & 7 -- 8.5 & 9 -- 11 & 11.5 -- 12.5 & 22 -- 26 & 25 -- 30 & 28 -- 33 & 30 -- 36\\
 \citet{marley-et-al-2007}$^a$ & 3 -- 5 & 4 -- 7 & 6 -- 10 & & & & & &\\
 \citet{chabrier-et-al-2000} & & & & & & 21 -- 26 & & 30 -- 35 &\\
 \citet{baraffe-et-al-2003} & & 4 -- 5 & 6 -- 7 & 9 -- 10 & & 21 -- 26 & & 30 -- 35 &\\
 \citet{baraffe-et-al-2008}$^b$ & & & $\sim$ 7  & $\sim$ 9 & & & & &\\
\hline
 \citet{baraffe-et-al-2003}$^c$ & & $\sim$ 5.5 & $\sim$ 8.5 & $\sim$ 10.5 & & $\sim$ 30 & & $\sim$ 38 &\\
\hline\hline
HR8799 c / d & \multicolumn{9}{c}{Luminosity $\log{L/L_{\odot}}= -4.7\,\pm\,0.1$ \citep{marois-et-al-2008}} \\
\hline
Model &  \multicolumn{9}{c}{Mass [M$_{\mathrm{Jup}}$] at age}\\
& 20 Myr & 30 Myr & 60 Myr & 100 Myr & 160 Myr & 590 Myr & 730 Myr & 1000 Myr &  1128 Myr  \\
\hline
 \citet{burrows-et-al-1997} & 6 -- 7.5 & 7.5 -- 9.5 & 11 -- 12 & 12.5 -- 13 & 13 -- 13.5 & 30 -- 38 & 35 -- 43 & 40 -- 48 & 41 -- 50\\
 \citet{marley-et-al-2007}$^a$ & 6 -- 8 & 8 -- 10 & & & & & & &\\
 \citet{chabrier-et-al-2000} & & 6 -- 7 & 8 -- 10 & 10 -- 11 & & 28 -- 34 & & 39 -- 46 &\\
 \citet{baraffe-et-al-2003} & & 6 -- 7 & 8 -- 10 & 10 -- 11 & & 27 -- 31 & & 37 -- 43 &\\
 \citet{baraffe-et-al-2008}$^b$ & & & $\sim$ 9 & & & & & &\\
\hline
 \citet{baraffe-et-al-2003}$^d$ & & $\sim$ 7.5 & $\sim$ 10.5 & $\sim$ 11.5 & & $\sim$ 38 & & $\sim$ 48 &\\
\hline
\end{tabular}
\begin{flushleft}
Remarks: All mass values were shortened to appropriate decimals and partly
rounded to halves or integers of a Jupiter mass. (a) Hot-start models. (b)
Non irradiated models. (c) Using M$_{Ks}$=14.05\,$\pm$\,0.08 mag
\citep{marois-et-al-2008}. (d) Using M$_{Ks}\sim$\,13.12 mag instead of
slightly different values M$_{Ks}$=13.13\,$\pm$\,0.08 mag for HR8799~c and M$_{Ks}$=13.11\,$\pm$\,0.12 mag
for HR8799~d, given in \citet{marois-et-al-2008}.
\end{flushleft}
\end{table*}

Using the model by \citet{baraffe-et-al-2003} and assuming the age range of $30$--$160$~Myr,
\citet{marois-et-al-2008} estimated the masses of companions to be
$7 M_{\mathrm{Jup}}$ ($5$--$11 M_{\mathrm{Jup}}$) for HR8799~b
and $10 M_{\mathrm{Jup}}$ ($7$--$13 M_{\mathrm{Jup}}$)
for HR 8799 c \& d.
We have recalculated possible masses of the three companions
with the aid of several state-of-the-art hot-start evolutionary models
from system's age and companions' luminosity.
In so doing, we allowed a broader
range of possible ages, as discussed in section 2.1.
Table~\ref{tab_masses_from_models} shows the results.
Note that some models give masses only for limited age and/or
mass ranges;  for instance, the model by \citet{marley-et-al-2007}
does not go beyond $10 M_{\mathrm{Jup}}$.
This explains why some positions in the table are not filled.

On the whole, the mass estimates we obtain are similar to those by \citet{marois-et-al-2008}.
The ages above $\sim 160$~Myr would lead to ``non-planetary'' masses above the standard
deuterium burning limit of $13.6 M_{\mathrm{Jup}}$.
On the other hand, none of the models predict
masses below $3-5$ (HR8799~b) and $6-8$ (HR8799~c,~d) Jupiter masses
even for the youngest plausible ages of $\sim 20$~Myr.

\subsection{Masses from stability requirement}

\begin{figure}
  \begin{center}
  \includegraphics[width=0.48\textwidth]{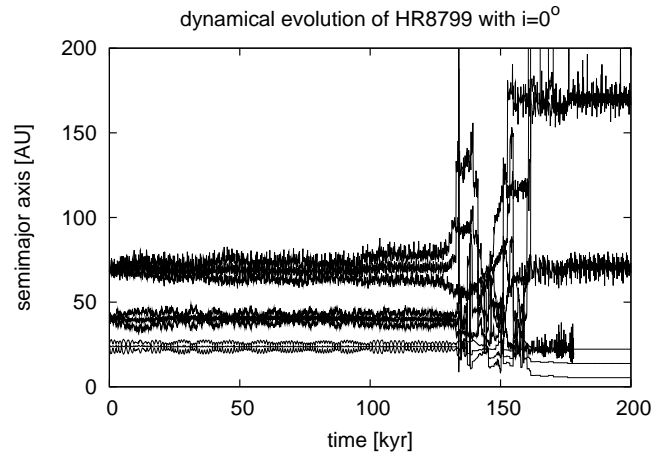}
  \end{center}
  \caption{
  Instability of the planetary system with nominal masses in the non-inclined case.
  For each of the planets, three curves correspond to
  semimajor axes,  pericentric distance, and apocentric distance.
  }
  \label{fig: hr8799_000_000}
\end{figure}

The simplest possible assumption one can make
is that the system is seen perfectly face-on (inclination $i=0$) and that all three planets
are initially in circular, co-planar orbits. However, as
\citet{fabrycky-murrayclay-2009} pointed out, such a system with masses reported
in \citet{marois-et-al-2008} would be unstable. This is readily
confirmed by our numerical integrations that are described below.
Fig.~\ref{fig: hr8799_000_000} shows the time evolution of the planetary
orbits; already after 134\,kyr a close encounter
between HR8799~c and d occurs and triggers instability of all three planets
ending with the ejection of HR8799~c 50\,kyr later. These timescales are
fully consistent with those reported by  \citet{fabrycky-murrayclay-2009}.

\begin{figure}
  \begin{center}
  \includegraphics[width=0.48\textwidth]{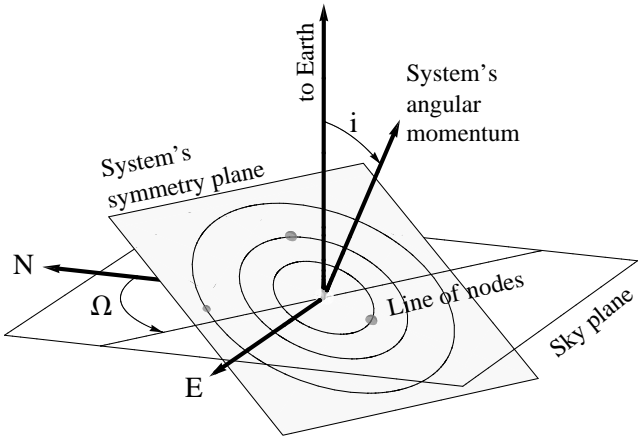}
  \end{center}
  \caption{
  Orientation of the system with respect to the line of sight.
  }
  \label{fig: orientation sketch}
\end{figure}

The studies by \citet{fabrycky-murrayclay-2009} and \citet{gozdziewski-migaszewski-2009}
were largely based on fitting simultaneously the observed positions and
differential proper motion of the companions
and checking the stability of the resulting systems in the course of their dynamical
evolution.
Noting that constraints on eccentricities from the differential proper motion
are rather weak, but clear indications exist for inclined orbits from the rotational
period analysis (see section \ref{subsec: Rotational period and inclination}),
here we employ a different method.
We confine our simulations to initially circular orbits
(and let the eccentricities evolve to non-zero values at later times),
but allow the symmetry plane
to have all conceivable non-zero inclinations and
an arbitrary orientation.
Thus, for the subsequent analysis we introduce
two angles (Fig.~\ref{fig: orientation sketch}).
One is the inclination $i$ itself,
measured between the angular momentum vector and a vector pointing toward
the observer. Another angle is the longitude of node $\Omega$
of the system's symmetry plane on the plane of the sky,
which is measured from north in the eastern direction.
We vary the inclination $i$ from $0^\circ$ to $45^\circ$,
thus extending the range suggested by section 2.2 to higher values.
This may be useful to accommodate a possible tilt of planetary orbits
to the stellar equator.
The rotation angle $\Omega$ is unconstrained by the observations.
It is sufficient to vary it from $0^\circ$ to $180^\circ$,
since the true mutual positions of all three planets
at ($i$, $\Omega$) and ($i$, $\Omega + 180^\circ$) would be exactly the same.
For each ($i$, $\Omega$)-pair we can convert the observed (projected)
instantaneous positions (astrocentric distances and positional angles) of the three planets
into their true positions in space.
If, further, initially circular orbits are assumed, the calculated distances
of planets will coincide with their initial semimajor axes.
We thus consider a two-parametric ($i$, $\Omega$) set of possible systems;
for each of them, the initial orbital configuration is fully and uniquely defined.

\begin{figure}
  \begin{center}
  \includegraphics[width=0.48\textwidth]{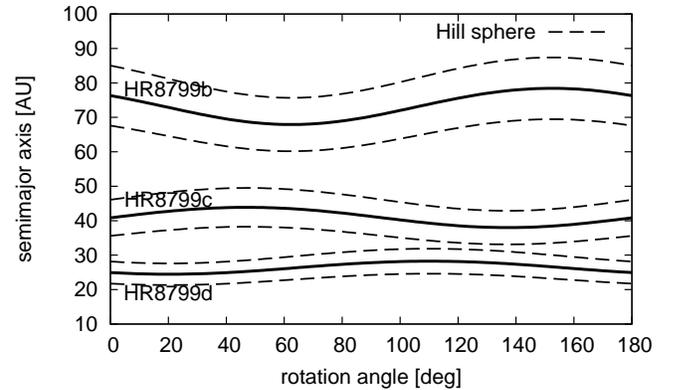}
  \end{center}
  \caption{
  Semimajor axes of HR8799b, c and d (solid lines) and their Hill radii for nominal masses (dashed)
  as a function of $\Omega$ for $i = 30^\circ$.
  }
  \label{fig_tilt_hill}
\end{figure}

\begin{figure}
  \begin{center}
  \includegraphics[width=0.48\textwidth]{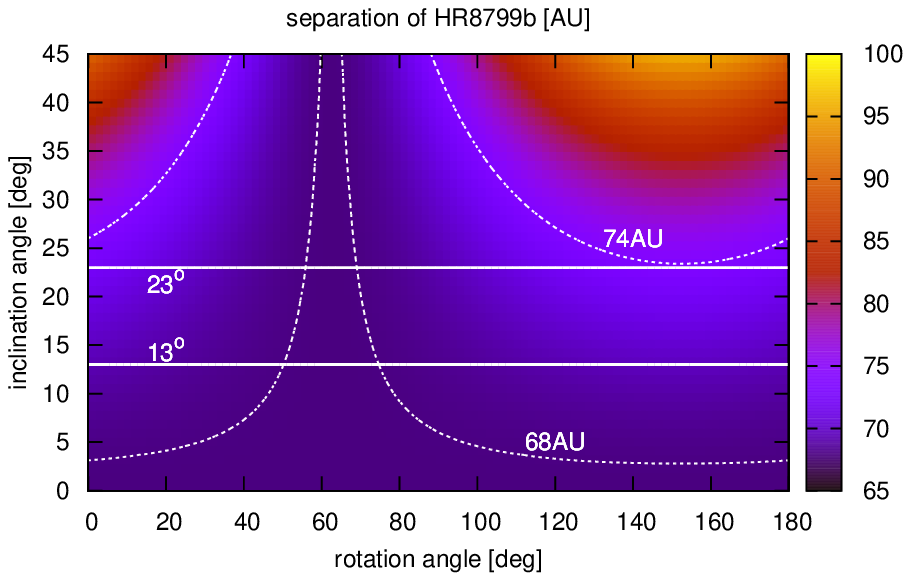}\\
  \includegraphics[width=0.48\textwidth]{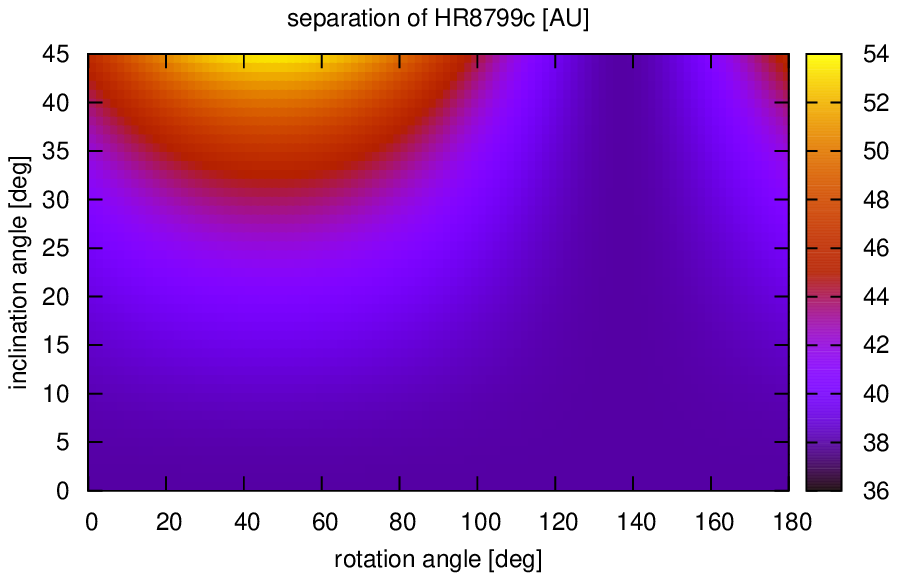}\\
  \includegraphics[width=0.48\textwidth]{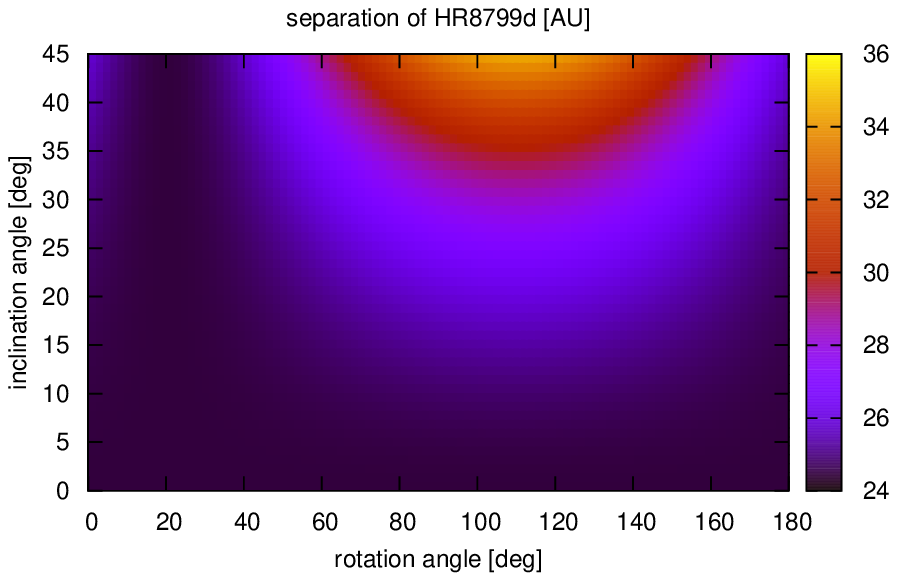}
  \end{center}
  \caption{Initial semimajor axes of HR8799b (\emph{top}), c (\emph{middle}) and d (\emph{bottom}) as a
  function of $i$ and $\Omega$.
  Solid horizontal lines in the uppermost panel border the range of
  inclination range of $i =  13$--$23^\circ$ for HR8799~b,
  reported by \citet{lafreniere-et-al-2009}.
  Dashed lines do the same for the semimajor axis range, $a = 68$--$74\AU$.
  }
  \label{fig_axes}
\end{figure}

\begin{figure}
  \begin{center}
  \includegraphics[width=0.48\textwidth]{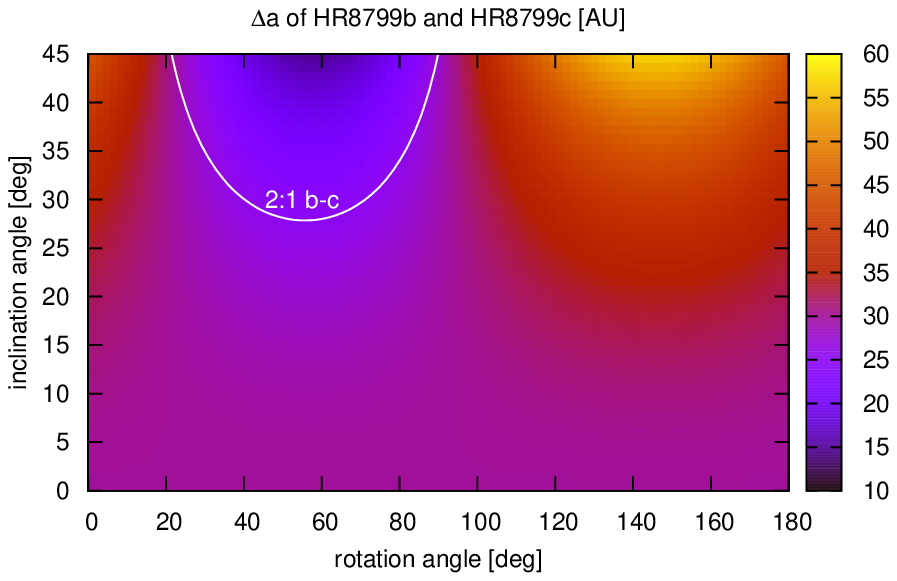}\\
  \includegraphics[width=0.48\textwidth]{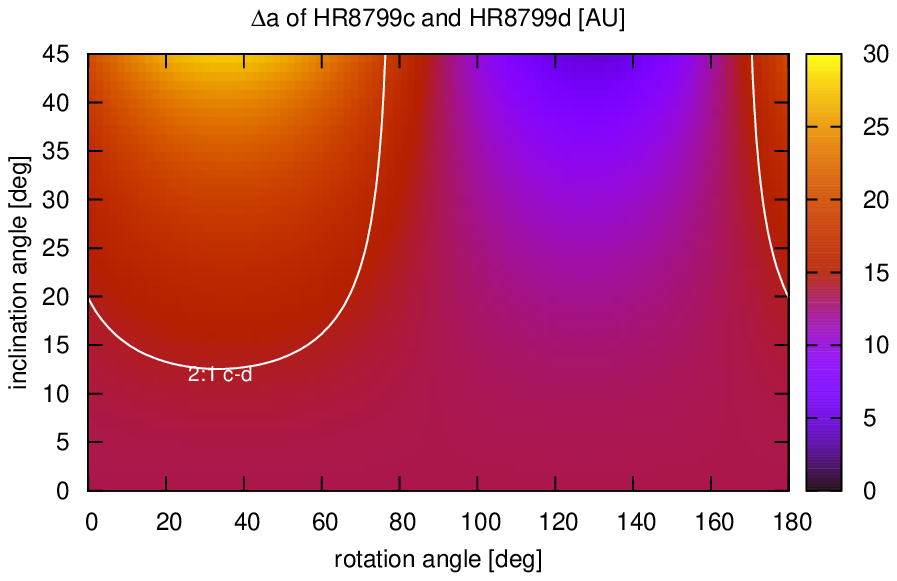}
  \end{center}
  \caption{Difference of initial semimajor axes $\Delta a$ between HR8799~b \& c (\emph{top}) and HR8799~c \& d (\emph{bottom}) as a function of $i$ and $\Omega$, as well as the location of the 2:1 commensurability of the periods for circular orbits.
  }
  \label{fig_delta_axes}
\end{figure}

Fig.~\ref{fig_tilt_hill} shows the astrocentric semimajor axes of the three planets
with an inclination of $i=30^\circ$ as a function of rotation angle $\Omega$,
as well as the corresponding Hill spheres for nominal planetary masses.
Similarly, Fig.~\ref{fig_axes} depicts the initial semimajor axis of the planets
depending on $i$ and $\Omega$.
Note that the recent analysis of HR8799~b by \citet{lafreniere-et-al-2009},
which is based on the archival HST/NICMOS data from 1998, yielded
a semimajor axis of $a\sim 68$--$74\AU$ and an inclination of $i \sim 13^\circ$--$23^\circ$.
Finally, Fig.~\ref{fig_delta_axes} plots the
difference of initial semimajor axes of HR8799~b and c, as well as HR8799~c
and d, again depending on $i$ and $\Omega$. From all these figures, it is
clearly seen that the orbital spacing, and therefore the stability,
might indeed strongly depend on the orientation of the system.

This expectation is fully confirmed by the main bulk of
numerical integrations that we performed with the aid
of the MERCURY6 package \citep{chambers-1999}.
We used the hybrid symplectic/Bulirsch-Stoer integrator with an adaptive
stepsize and a $10^{-14}$ angular momentum conservation accuracy,
which changes to Bulirsch-Stoer algorithm at distances less than 3
Hill radii. Output has been stored every 1000 years. Each integration
terminated when two planets had a distance less than half the Hill radius or
after an integration time of $t_\mathrm{max} = 100$\,Myr. In all cases we
assumed  a stellar mass of $1.5M_\odot$ and a distance of $39.4$\,pc to
convert the separation angle into the projected astrocentric distance.
The three planets started at positions at the epoch of 2008 Sept. 18
\citep[see Table 1 in][]{marois-et-al-2008}.
We used 3 different sets of possible planet masses (see Table \ref{tab: planetmasses})
and checked stability for
$i \in \{0, 1,\dots, 45^{\circ}\}$ and $\Omega \in \{0, 5, \dots, 180^{\circ}\}$.
Note that, although we assumed initially circular orbits, it does not
mean that the orbits stay circular at later times.
Just the opposite: the mutual perturbations always force eccentricities
to take values in the range between zero and approximately 0.1, so that the
initial circularity is ``forgotten'' by the system.

\begin{table}[htb!]
  \centering
  \caption{Three sets of planet masses used in numerical integrations}
  \label{tab: planetmasses}
  \begin{tabular}{lccc} \toprule
    \midrule
    HR8799        & b & c & d \\
    \midrule
    low mass      & $5M_\mathrm{Jup}$  & $ 7M_\mathrm{Jup}$ & $7M_\mathrm{Jup}$ \\
    nominal mass  & $7M_\mathrm{Jup}$  & $10M_\mathrm{Jup}$ & $10M_\mathrm{Jup}$ \\
    high mass     & $11M_\mathrm{Jup}$ & $13M_\mathrm{Jup}$ & $13M_\mathrm{Jup}$ \\
    \midrule
    \bottomrule
  \end{tabular}
\end{table}

\begin{figure}
  \begin{center}
  \includegraphics[width=0.48\textwidth]{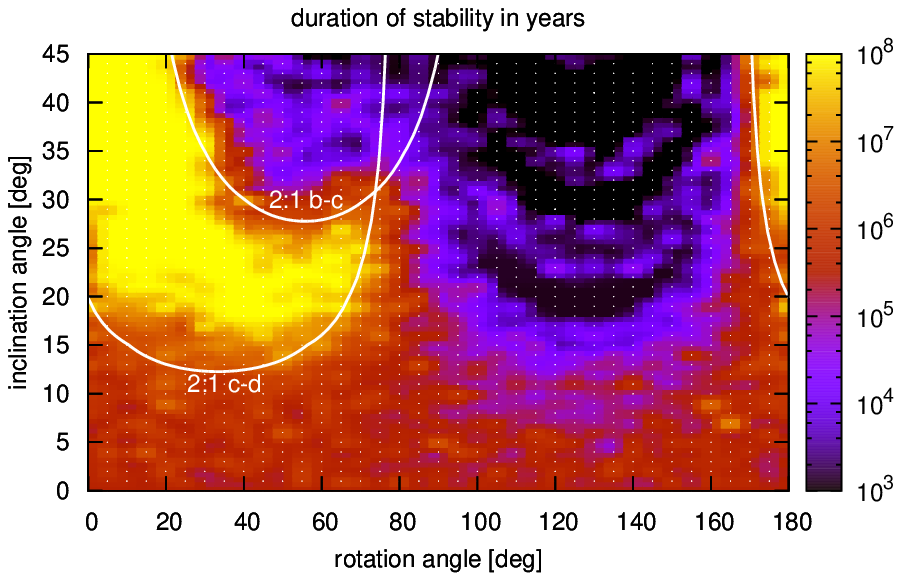}
  \includegraphics[width=0.48\textwidth]{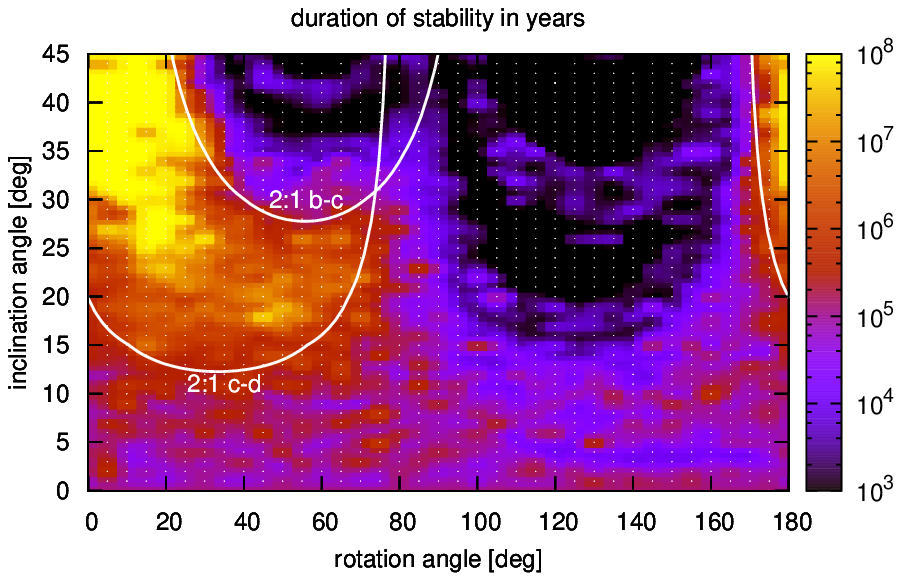}
  \includegraphics[width=0.48\textwidth]{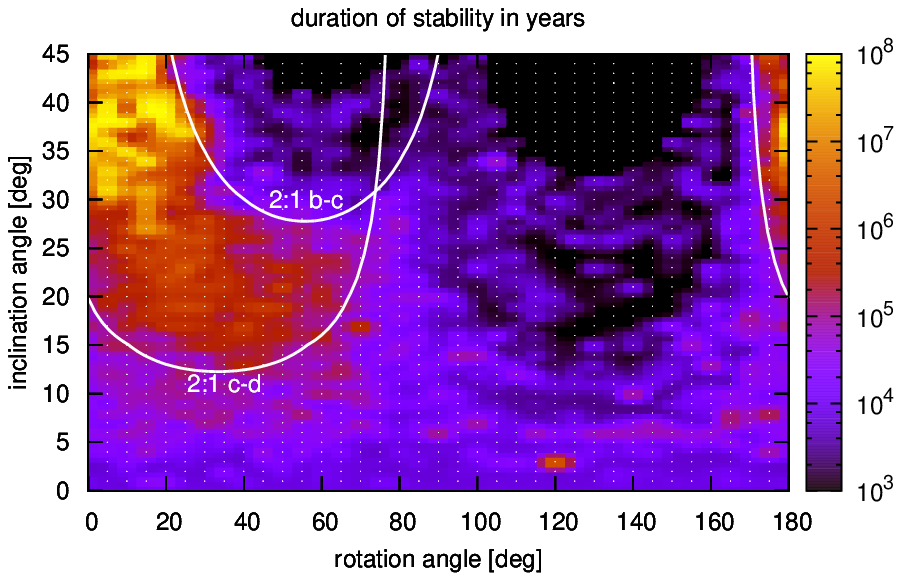}
  \end{center}
  \caption{Duration of stability depending on $i$ and $\Omega$ for different sets of planetary masses
  given in  Tab.~\ref{tab: planetmasses}:
  \emph{top:} low, \emph{middle:} nominal, \emph{bottom}: high.
  A grid of dots corresponds to the actual set of numerical runs (one dot = one run).
  Curves show where the periods between b-c and c-d would have a commensurability of 2:1 if the
  orbits were exactly circular.
  }
  \label{fig: stability_log}
\end{figure}

The results of these integrations are presented in Fig.~\ref{fig: stability_log}.
It depicts the time interval until the first close encounter~--- as a proxy for stability~---
with an upper limit of 100\,Myr.
It is seen that all three planets may be stable for 100~Myr for either set of planetary masses,
but only for some of all possible orientations. Specifically, an
inclination of $\ga 20^{\circ}$ is required and the rotation angle $\Omega$
must lie within the range from $\approx 0^\circ$ to $\approx 50^\circ$.
The higher the masses, the narrower the ``stability spot'' in the $(i, \Omega)$-plane.
Further, we conclude that it is the inner pair (c and d) that tends to destroy the
stability.
Indeed, comparing Fig.~\ref{fig: stability_log} to
Fig.~\ref{fig_delta_axes} one sees that the most stable regions are
those where $\Delta a$ of HR8799~c and d is the largest,
whereas that of HR8799~b and c is not.
Finally, a comparison of Fig.~\ref{fig: stability_log} and
Fig.~\ref{fig_axes} shows that the position of the ``stability spot'' in
Fig.~\ref{fig: stability_log}
roughly matches the inclination of $i=13$--$23^\circ$
and the semimajor axis of $a = 68$--$74\AU$ of the outermost planet
reported by \citet{lafreniere-et-al-2009}.

\begin{figure}
  \begin{center}
  \includegraphics[width=0.48\textwidth]{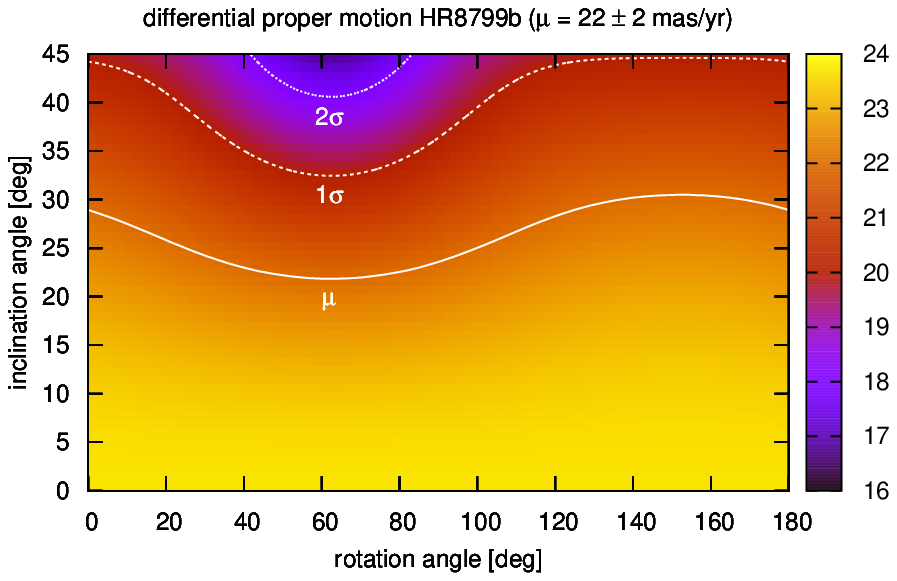}\\
  \includegraphics[width=0.48\textwidth]{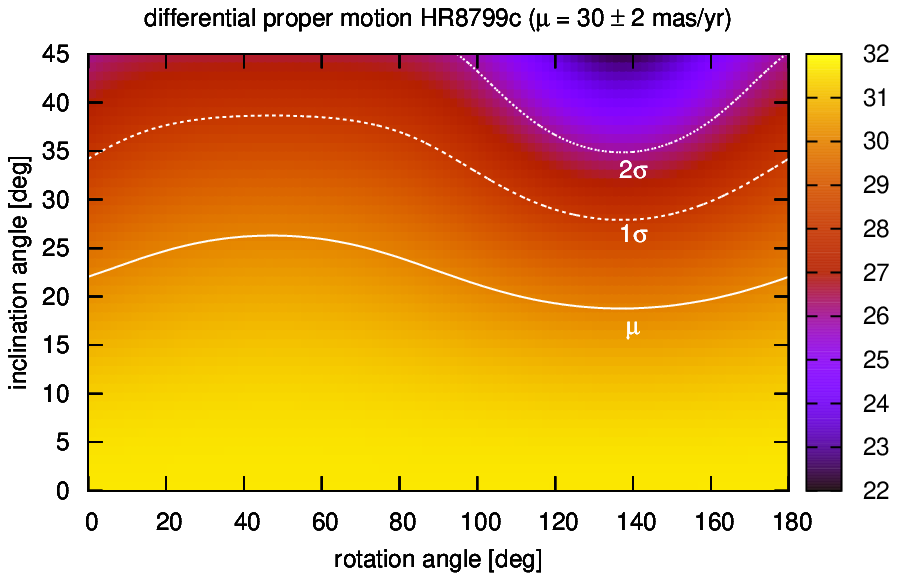}
  \end{center}
  \caption{
  Sky-projected differential proper motion that HR8799~b (\emph{top}) and c (\emph{bottom})
  would have for each pair of $i$ and $\Omega$ (assuming circular orbits).
  The isolines are at $\mu$, $\mu - \sigma$, and $\mu - 2\sigma$ ($\sigma = 2$~mas/yr), where
  $\mu$ is the differential proper motion derived from the measurements, as described in the text.
  Note that the $\mu + \sigma$ and $\mu + 2\sigma$ curves for both companions fall outside
  the plots.
  A panel for HR8799~d is not included because the uncertainty
  ($\mu = 42$~mas/yr $\pm 27$~mas/yr) is too large.
  }
  \label{fig_orbital_motion}
\end{figure}

We now check whether all considered geometries are consistent with the measured
differential proper motion of the companions.
Fig.~\ref{fig_orbital_motion} depicts
the projected differential proper motions $\mu$ that the planets in circular orbits would have
for each pair of $i$ and $\Omega$.
Overplotted are the values of $\mu$ actually measured together with
their $1\sigma$ and $2\sigma$ deviations.
For HR8799~c, we used $\mu = 30 \pm 2$\,mas/yr from \citet{marois-et-al-2008}.
In contrast, for HR8799~b we derived the differential proper motion of
$22 \pm 2$\,mas/yr by combining the Marois et al. measurements with
the 1998 data  from \citet{lafreniere-et-al-2009}.
Note that further adding the 2002 positions from \citet{fukagawa-et-al-2009}
does not change this result.
For comparison, the differential proper motion of HR8799~b
given in \citet{marois-et-al-2008} is $25 \pm 2$\,mas/yr.
In the non-inclined case, the calculated differential proper motion
is always within $1\sigma$ of the measured one.
For HR8799~b, all orbits with an inclination $< 33^{\circ}$ lie within
$1\sigma$ of the measured value. For HR8799~c, the same is true for $i < 28^\circ$.
Nearly the whole parameter range of $(i, \Omega)$ explored here is compatible
with observations within $2\sigma$. It is easy to show that taking into account
actual low eccentricities up to $\approx 0.1$ acquired by the planets would
not change this conclusion.

\begin{figure}
  \begin{center}
  \includegraphics[width=0.48\textwidth]{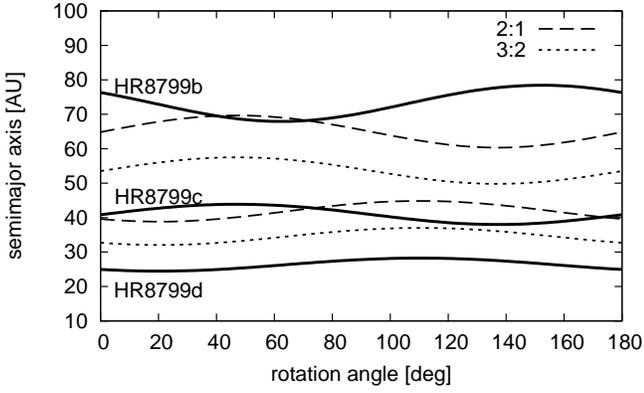}
  \end{center}
  \caption{
  Semimajor axes of HR8799b, c and d (solid lines) and positions of
  external 2:1 (dashed) and 3:2 nominal resonances (dotted) of two inner planets as a
  function of $\Omega$ for $i = 30^\circ$.
  }
  \label{fig_tilt_res}
\end{figure}

To understand the stability region shown in Fig.~\ref{fig: stability_log},
we check the possibility that it may be related to resonances.
From Fig.~\ref{fig_tilt_res} it is obvious that the two inner planets are
close to the 2:1 resonance.
With still coplanar and circular orbits, we looked for combinations of $i$, $\Omega$
that would correspond to the 2:1 commensurability.
The resulting loci of the nominal 2:1 resonance in the
($i$,$\Omega$)-plane (Fig.~\ref{fig: stability_log})
just encircle the stability region. This strengthens the hypothesis that the
stability may be directly related to the 2:1 resonance.

We then checked whether or not the orbits of HR8799~c and d within the stability region
are indeed locked in the resonance. To this end, we have calculated the resonant argument
$\varphi_\mathrm{cd} = 2 \lambda_\mathrm{c} - \lambda_\mathrm{d} - \omega_\mathrm{d}$,
where $\lambda_c$ and $\lambda_d$ are the mean longitudes of HR8799~c and d
and $\omega_d$ is the argument of pericenter of the latter planet.
We found that {\em all} stable orbits are indeed resonant.
Interestingly, the initial values of $\lambda_c$ and $\lambda_d$ adopted in all numerical runs
were such that the planets are not locked in the resonance initially, but~--- in all stable cases~--
the system swiftly ``slips'' into the resonance, which makes it safe.
The resonant argument $\varphi_\mathrm{cd}$ librates around $0^\circ$ with an amplitude
(which we calculated as a standard deviation) of $22^\circ$--$100^\circ$.
For comparison, the non-resonant case would have a standard deviation of $\sim 103.9^\circ$.
However, we made sure that even the cases with the libration amplitude
up to $100^\circ$ are resonanceces, albeit shallow.
In these cases, the resonant argument circulates rather than librates.
Thus the phase trajectories on the $e_c \cos\varphi_\mathrm{cd}$--$e_c \sin\varphi_\mathrm{cd}$
plane are circles with an offset from (0,0), which is indicative of a resonant locking.

\begin{figure}
  \begin{center}

\framebox[0.24\textwidth]{
\begin{minipage}{0.24\textwidth}
  \centering
  $(i, \Omega) = (25^\circ, 15^\circ)$
  \includegraphics[width=0.99\textwidth]{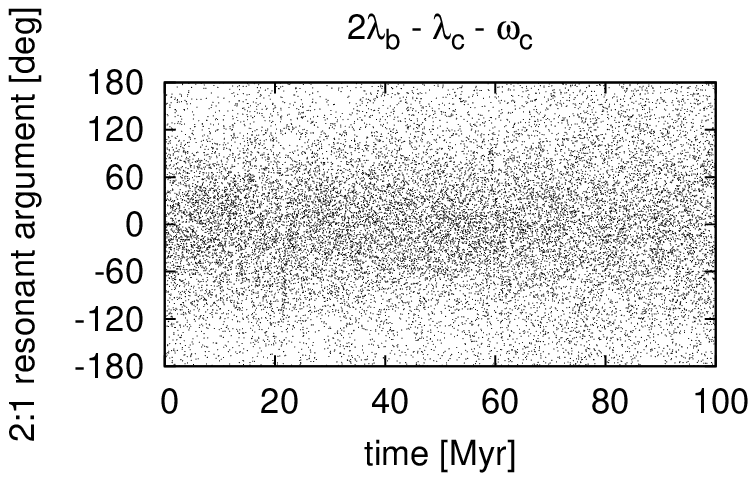} \\
  \includegraphics[width=0.99\textwidth]{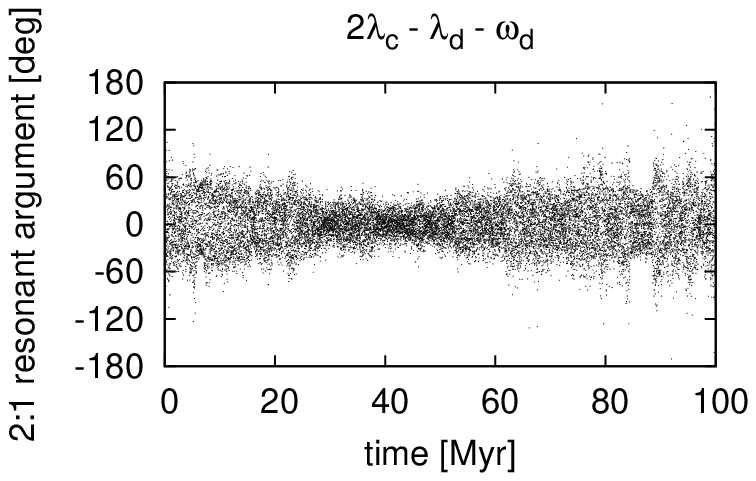}

\end{minipage}
}
\framebox[0.24\textwidth]{
\begin{minipage}{0.24\textwidth}
  \centering
  $(i, \Omega) = (29^\circ,20^\circ)$
  \includegraphics[width=0.99\textwidth]{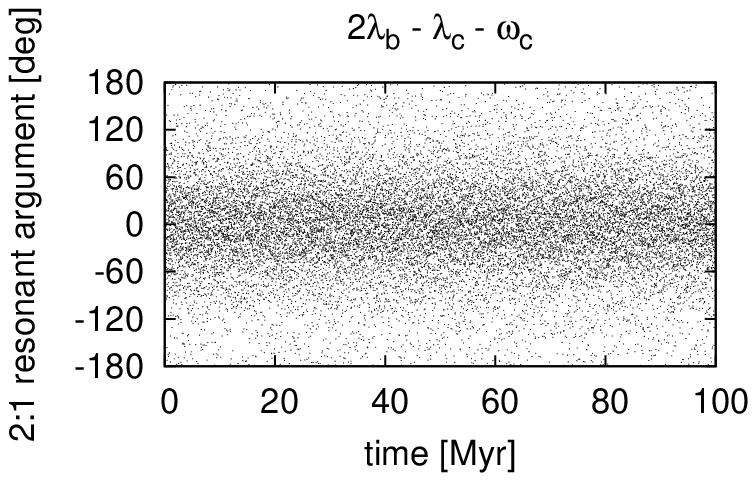} \\
  \includegraphics[width=0.99\textwidth]{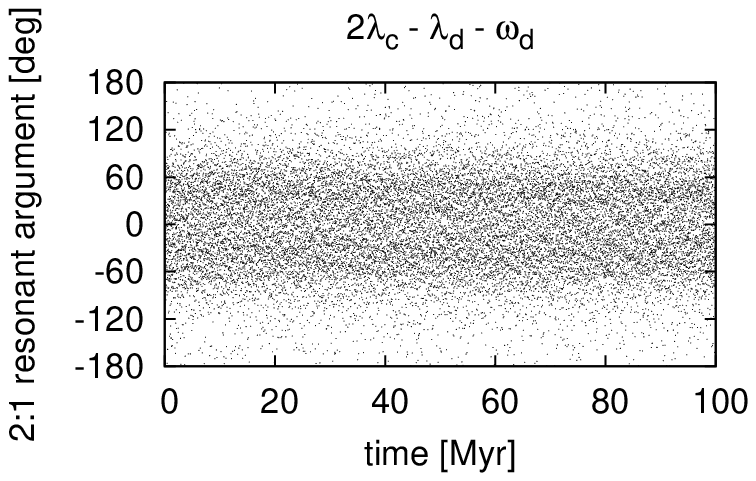}
\end{minipage}
}
  \end{center}
  \caption{Typical behavior of resonant arguments for the 2:1 mean motion resonance
  between HR8799~b \& c (\emph{top}) and HR8799~c \& d (\emph{bottom}).
  Nominal planetary masses are assumed.
  The libration amplitude is $86^\circ$ (\emph{left top}),
                                  $36^\circ$ (\emph{left bottom}),
                                  $63^\circ$ (\emph{right top}), and
                                  $60^\circ$ (\emph{right bottom}).
  }
  \label{fig: 2to1_resoarg}
\end{figure}

For all stable configurations we then calculated the resonant argument for the two outer planets,
HR8799~b and c, defined as
$ \varphi_\mathrm{bc} = 2 \lambda_\mathrm{b} - \lambda_\mathrm{c} - \omega_\mathrm{c}$.
We found that they are in a 2:1 resonance, too,
with the standard deviation in the range $36^\circ$--$97^\circ$.
Note that in all stable cases at least one of the resonances, c-d or b-d,
is strong, as suggested by a low libration amplitude.
Two typical examples of the time evolution of the resonant argument for both planetary pairs
are shown in Fig. \ref{fig: 2to1_resoarg}.
Thus our results are consistent with those by
\citet{fabrycky-murrayclay-2009} and \citet{gozdziewski-migaszewski-2009}, who suggested a
double resonance 4:2:1 as the likely ``survival recipe'' for the entire three-planet system.

As noted above, the major danger of system's destabilization comes from the
two inner planets which are likely more massive and more tightly spaced.
Keeping this in mind and taking into account that data on the innermost companion
(e.g. its differential proper motion) are as yet the least reliable, we checked the dynamical
stability properties the system would have without HR8799~d.
We used the same setup for MERCURY6 and restricted our analysis
to three cases: non-inclined configuration, $i = 15^\circ$, and $i = 30^\circ$.
The overall result is that, as expected, the absence of the inner companion
would drastically improve stability:
\begin{itemize}
\item
{\em The non-inclined configuration} becomes
stable over a period of 100\,Myr~--- not only for the nominal masses, but also
for much higher masses up to $M_b = 22 M_{\mathrm{Jup}}$ and $M_c = 30 M_{\mathrm{Jup}}$.
The rapid breakdown of the system in less than 10~kyr would only be guaranteed
with masses as high as $M_b = 33 M_{\mathrm{Jup}}$ and $M_c = 45 M_{\mathrm{Jup}}$;
\item
{\em For $i = 15^\circ$} and all $\Omega = 0^\circ$--$180^\circ$,
the system with nominal masses is always stable over 100~Myr;
\item
{\em For $i = 30^\circ$},
the system with nominal masses is unstable
for $\Omega= 40^\circ$--$75^\circ$ and stable otherwise.
\end{itemize}

\section{Dynamical interaction between the planetesimal belts and the planets}

To check where the outer and inner planetesimal belts (maintaining
dust rings d and b) would be truncated
by planets, we have picked up one exemplary angle configuration
from within the ``stability spot'' seen in Fig.~\ref{fig: stability_log}.
The point chosen is $(i, \Omega) = (30^{\circ}, 10^{\circ})$, which
implies initial semimajor axes of 74.7, 41.8 and 24.6~AU for HR8799~b, c and d,
respectively.

\begin{figure*}
  \begin{center}
  \includegraphics[width=0.45\textwidth]{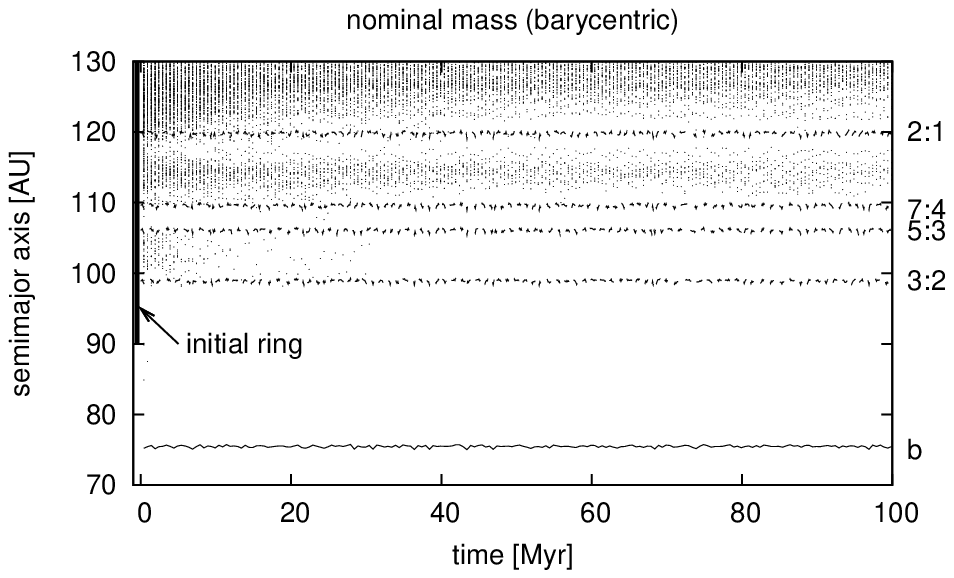}
  \includegraphics[width=0.45\textwidth]{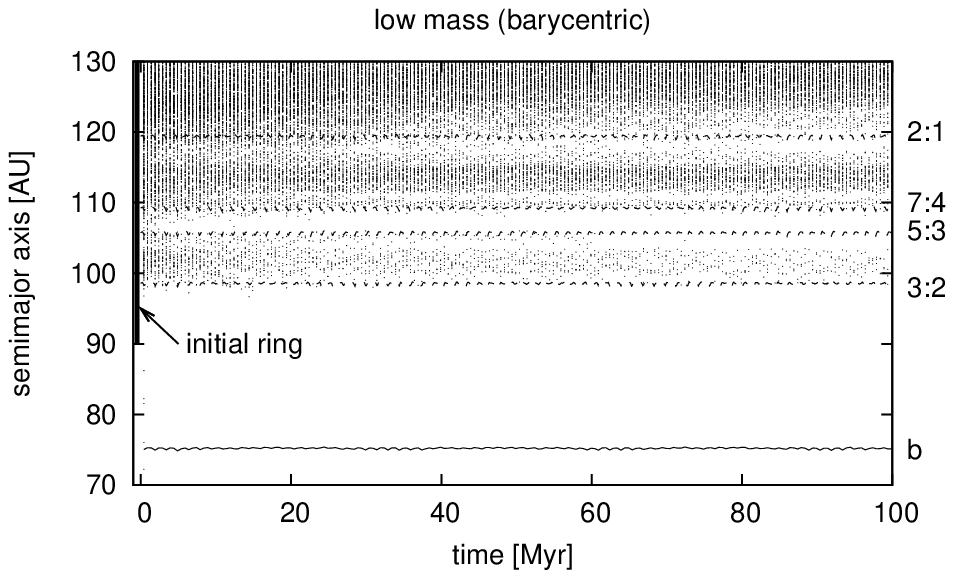} \\
  \includegraphics[width=0.45\textwidth]{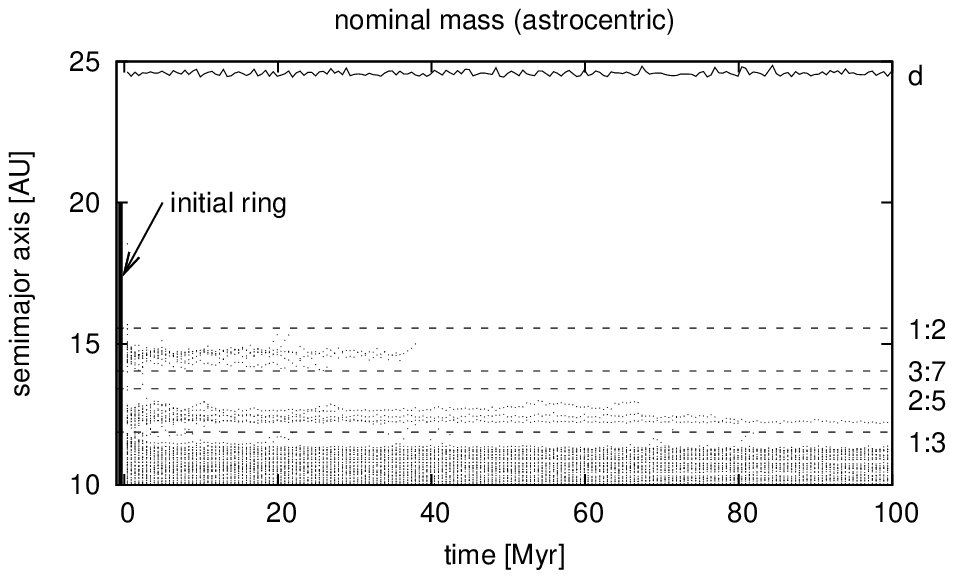}
  \includegraphics[width=0.45\textwidth]{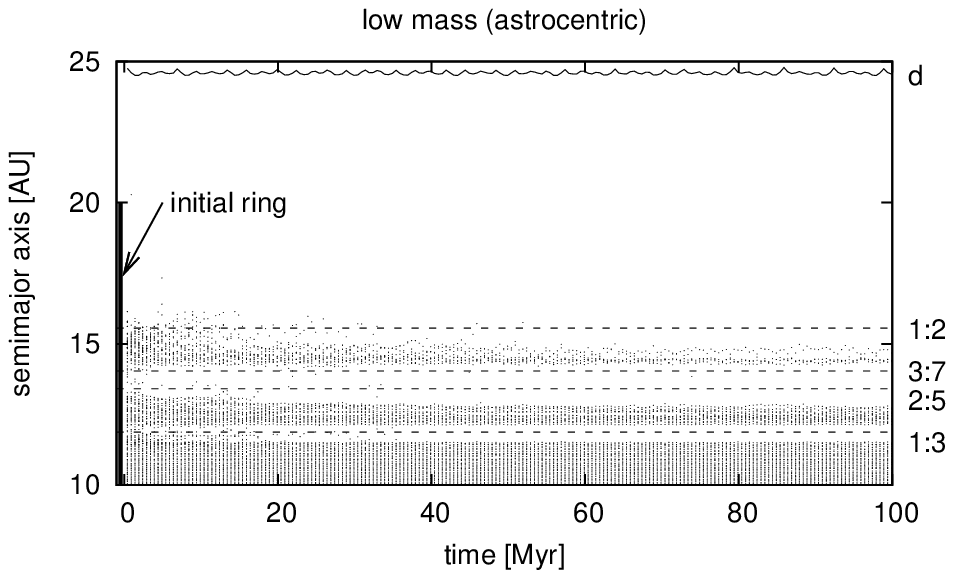}
  \end{center}
  \caption{
  Evolution of semimajor axes of planetesimals over 100\,Myr of
  for the nominal (\emph{left}) and low (\emph{right}) planet masses.
  Starting planetesimal rings are shown with vertical bars.
  Solid and dashed lines represent the location of planets and important
  resonances, respectively.
  Note that the outer ring (\emph{top}) is plotted in barycentric osculating elements and the inner ring
  in astrocentric ones. This choice leads to a sharper visibility of the resonance positions.
  }
  \label{fig: rings}
\end{figure*}

\begin{figure}
  \begin{center}
  \includegraphics[width=0.45\textwidth]{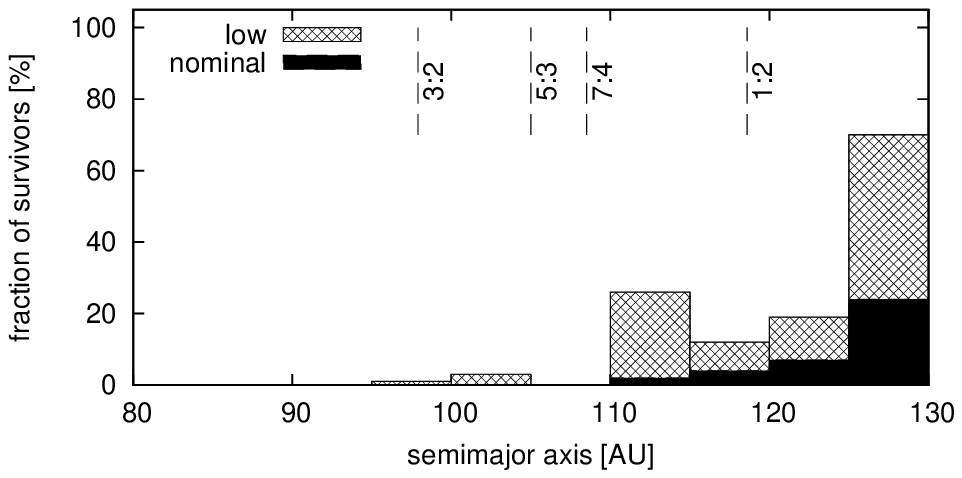}\\
  \includegraphics[width=0.45\textwidth]{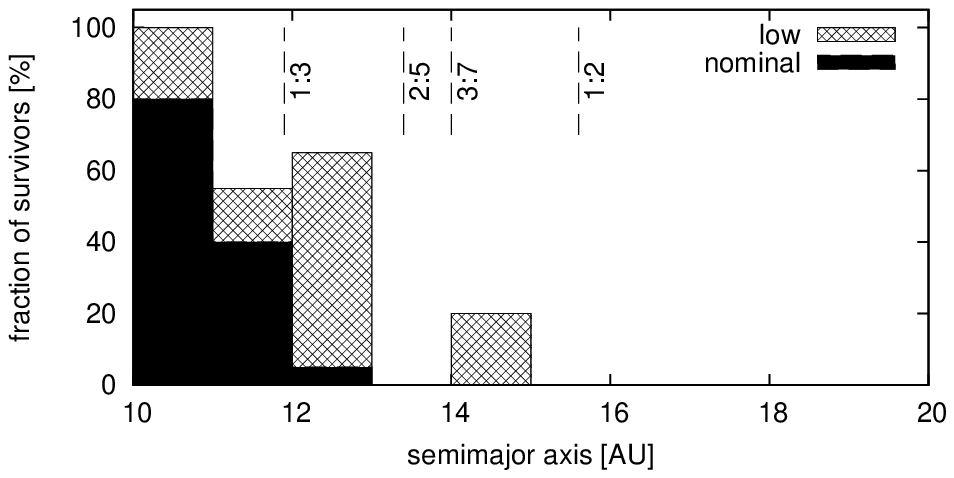}
  \end{center}
  \caption{
  Fraction of left-over planetesimals after 100~Myr of evolution in the outer (\emph{top})
  and inner (\emph{bottom}) rings.
  Filled/hatched bars: nominal/low planetary mass.
  }
  \label{fig: survivors}
\end{figure}

We launched 1000 massless planetesimals with
uniformly distributed orbital elements
($e_{p} = 0$--$0.2$, $i_{p} = 0$--$10^{\circ}$, $\{\omega_{p}, \Omega_{p}, M_{p}\} = 0$--$360^\circ$).
Of these, 200 planetesimals were initially confined to a ring between
$a_{\mathrm{inner}} = 10$--$20\AU$
and the other 800 between $a_{\mathrm{outer}} = 90 - 130\AU$.
We integrated their orbits over 100\,Myr with an accuracy of $10^{-12}$ for low
and nominal planetary masses, as given in Table \ref{tab: planetmasses}.
The high mass case was excluded, because
the stability of companions themselves at $(i, \Omega) = (30^{\circ}, 10^{\circ})$
was only marginal.

Figure~\ref{fig: rings} shows that both the inner part
of the outer ring and the outer part of the inner ring are swiftly cleared by
the adjacent planet (b and d, respectively). The belt of remaining planetesimals
develops ``Kirkwood gaps'' at the positions of major resonances. Over tens of Myr,
the gaps get progressively more pronounced. Simultaneously,
the fraction of particles surviving between them gradually decreases.
As expected, the survival probability of planetesimals at a given distance
is larger for lower masses of the companions.

For both rings and for the nominal- and low-mass cases,
Fig.~\ref{fig: survivors} presents the fraction of planetesimals
that survived after 100~Myr in orbits with different initial semimajor axes.
For nominal planetary masses, the outer ring shows a considerable depletion,
with only $\sim 10$ to 20\% planetesimals surviving even outside $120\AU$.
There are almost no survivors inside the 5:3 and 7:4 resonances with HR8799~b at
$\approx 105$--$110\AU$.
In the low-mass case the survival fraction in the outer ring is appreciably higher
(15--70\% between $\approx 110$--$130\AU$).
The inner ring retains 80--100\% of planetesimals inside 11~AU.
The outer edge of the ring is at 13~AU (nominal-mass planet) to 15~AU (low-mass one).

These results have to be compared with dust locations found from the SED fitting
in section 3.2. As noted there, the far-infrared to millimeter part of the SED requires
dust in the outer ring as close as 120~AU from the star; we have just shown that a significant
fraction of planetesimals survives outside 120~AU after 100~Myr,
at least in the low-mass case.  In the nominal-mass case, the fraction of survivors
is lower, but any firm conclusions appear premature, since the location of the
outer ring is in fact not well-constrained (see section 3.2.).
Next, the IRS spectrum interpretation requires dust in the inner ring at
least at 10~AU away from the star.
This is comfortably within the stability zone inside the orbit of HR8799~d for
both the nominal-mass and low-mass cases.
Moreover, 10~AU quoted above is the distance where {\em dust} is required;
as discussed in section 3.2, the parent planetesimals would orbit closer to the star,
being yet safer against the perturbations of the innermost planet than their dust.
In summary, our analysis of the outer system might slightly favor the low-mass case,
but would not really pose any additional strong constraints to the planetary masses.

\section{Conclusions and discussion}

\subsection{Conclusions}

In this paper, we made an attempt of a coherent analysis of various portions of observational data
currently available for the system of a nearby A5 star HR8799, which hosts debris dust
as well as three planetary candidates recently discovered via direct imaging
\citep{marois-et-al-2008}.
A dedicated analysis of all known components of the system
(the central star, imaged companions, and dust)
leads us to a view of a complex circumstellar system
(Fig.~\ref{fig_system}).
It contains at least three planets in nearly-circular coplanar orbits bordered
by two dust-producing planetesimal belts, one outside the planetary region and another
inside it. Each planetesimal belt is encompassed by a dust disk. The outer dust disk
may have a considable extention, perhaps several hundreds of AU.

Our specific conclusions are as follows:

\begin{figure}
  \begin{center}
 \includegraphics[width=0.48\textwidth]{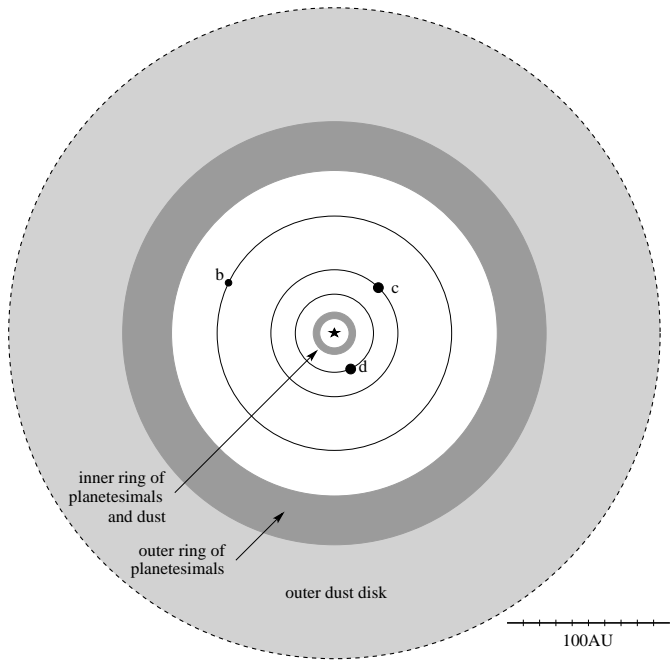}
  \end{center}
  \caption{
  A schematic view of the system HR8799.
  }
  \label{fig_system}
\end{figure}

\begin{enumerate}

\item
With previous estimates of  stellar age ranging from $\approx 20$  to $\approx 1100$~Myr,
high luminosity of observed cold dust may favor younger ages of $\la 50$~Myr.
A younger age would automatically lower the masses for all three companions,
estimated through evolutionary models, making the masses more consistent
with dynamical stability results (see conclusion 5).

\item
The system is seen nearly pole-on.
Our analysis of the stellar rotational velocity suggests
an inclination of $13$ -- $30^{\circ}$, whereas $i \ga 20^\circ$ seems to be mandatory
for the system to be dynamically stable (see conclusion 5).
Thus we arrive at a probable inclination range of $20$ -- $30^{\circ}$.

\item
Our analysis of the available mid-infrared to millimeter photometry and spectrophotometry
data reveals the presence of two dust rings,
and therefore two parent planetesimal belts, an ``asteroid belt'' at $\sim 10$~AU and a ``Kuiper
belt'' at $\ga 100$~AU.
The dust masses are estimated to be $\approx 1 \times 10^{-5}~M_\oplus$ and
$4 \times 10^{-2}~M_\oplus$ for the inner and outer ring, respectively.

\item
Assuming that the system is indeed rather
young ($\la 50$~Myr) and based on the photometry
of the companions reported by \citet{marois-et-al-2008},
our estimates with several evolutionary models suggest
the masses of the companions to be lower than $7$ to $10 M_{\mathrm{Jup}}$.

\item
We show that all three planets may be stable in the mass range
suggested in the discovery paper by Marois et al. 2008 (between 5 and $13 M_{\mathrm{Jup}}$),
but only for some of all possible orientations.
For $(M_b,M_c,M_d) = (5,7,7) M_{\mathrm{Jup}}$, an inclination $i \ga 20^\circ$ is
required and the line of  nodes of the system's symmetry plane on the sky must lie within
$0^\circ$ to $50^\circ$ from north eastward.
For higher masses $(M_b,M_c,M_d)$ from $(7,10,10) M_{\mathrm{Jup}}$
to $(11,13,13) M_{\mathrm{Jup}}$, the constraints
on both angles are even more stringent.
The stability of the two inner planets is due to locking in the
2:1 mean-motion resonance, and the stability of the outer couple
is supported by the 2:1 commensurability, too.
However, in many stable cases only one of the two resonances is strong.
Another one is often shallow, with a circulating rather than librating
resonant argument.
For ``wrong'' orientations, the stability only seems possible with
planetary masses lower than most evolutionary models would predict
even for the youngest possible age (cf. Table~\ref{tab_masses_from_models}).
Should this be the case, this would necessitate revisions to
the models.

\item
Both dust/planetesimal belts appear to be dynamically stable against
planetary perturbations, provided the masses of companions are such
that they themselves are dynamically stable against mutual perturbations.
\end{enumerate}

\subsection{Prospects for future observations}

Given the paucity of observational data available to date, many of the
quantitative estimates listed above are quite uncertain and should
be taken with caution.
However, there is little doubt that new observations will arrive soon,
verifying these estimates and on any account reducing the uncertainties.

Firstly, new observations of the planets themselves
are expected and would be of great value.
Better astrometry, and therefore a
better determination of the orbits should  become possible
with improving instruments and methods of astrometric observations,
and because of the longer time spans.
Next, new photometry observations are needed.
The companions are detected so far in JHK and L$^\prime$.
The SED is relatively flat in this wavelength range
for objects with temperature of roughly 500 to 1500~K.
Imaging photometric detections of the companions in
either Gunn z ($1\mum$) or M ($5\mum$) must be possible
with 8 to 10m class telescopes.
They would allow one to constrain
the objects' temperature, because the colors z-J and L-M
and their differences depend on temperature strongly
over the relevant range of 500 to 1500 K.
Further, spatially resolved spectroscopy of the companions
may be possible with VLT/Sinfoni
or Subaru/IRSC, but would be very challenging.
If successful, it would place tighter constraints on
temperature and gravity and, hence, radius and mass of the companions.

Secondly, new data on the dust portion of the system
would be particularly promising.
For instance, a better mid-infrared photometry would
result in more reliable dust mass and location of the inner
dust belt (``exozodi'').
One could also think of near- and mid-IR interferometry observations, which have proven
very successful not only for exozodi studies, but also for stellar radius determination
\citep[see, e.g.,][]{difolco-et-al-2004,absil-et-al-2006,difolco-et-al-2007,absil-et-al-2008}.
While HR8799 is too faint to be observed with the presently operating CHARA/FLUOR and Keck
Interferometry Nuller instruments
(e.g. for CHARA the K-magnitude of $\la 4$\,mag is needed, whereas HR8799 has K = 5.2 mag),
this should become possible in the near future, for instance with VLTI/PIONIER and the LBTI Nuller.
More observational effort is required for the outer disk as well.
Resolving the outer debris disk,
especially in scattered light, would answer several key questions at a time.
On the one hand, it would further constrain the inclination of the entire system
and the orientation of its line of nodes on the sky plane, drastically
reducing the parameter space assumed in the dynamical simulations.
On the other hand, the precise location of the inner rim of the disk
could place a direct upper limit on the mass of HR8799~b, much in the same way
it was recently done for the Fomalhaut planet \citep{kalas-et-al-2008, chiang-et-al-2009}.
Since a high dustiness of the debris disk of HR8799 makes it
a relatively easy target, whereas advantages of resolving it are obvious,
success can be expected in the near
future\footnote{After submission of this paper, we became aware of recent
successful observations: the outer disk of HR8799 was resolved at $24$ and $70\mum$ with
Spitzer/MIPS (Kate Su, pers. comm).
The data analysis in ongoing.}.

Once the location and masses of the dust belts are better constrained from observations,
it will become possible to access the position, masses, and other properties of
directly invisible planetesimal belts that produce and sustain that dust.
This could be done with the help of elaborate collisional models
\citep{krivov-et-al-2008}. The results could provide additional clues to the
formation history of the system.

\subsection{Origin and status of the system}

The HR8799 system is among a few systems known to date to possess more than one planet
{\em and} at least one planetesimal/dust belt. Other examples are
HD38529 \citep{moromartin-et-al-2007b} and HD69830 \citep{beichman-et-al-2005b, lisse-et-al-2007}.
Further, it is not the only system with directly imaged companions whose masses
most likely fall into the ``planetary'' (as opposed to brown dwarf) range.
However, in some sense HR8799 does appear unique for the moment. The orbits of companions
extending up to $\approx 70$~AU are large, and their masses most likely too~--
probably almost at the limit of their dynamical stability against mutual perturbations.
Even though we strongly argue that the masses are well below the deuterium burning
limit, it is not clear whether the companions have formed in a ``planetary'' way
(from the protoplanetary disk) or ``stellar'' way (as a multiple stellar system).
In this sense, it remains questionable whether we are dealing with ``true''
planets. An argument in favor of the ``stellar'' formation would be, for instance,
a low metallicity \citep[Fe/H~$\approx -0.47$,][]{gray-kaye-1999}, atypical
of~--- although not the lowest among~--- the known planet host stars.
The low metallicity is particularly unusual for a system with several high-mass planets.
``Planetary'' way of formation in situ could be feasible through the gravitational instability
(GI) \citep{cameron-1978,boss-1998}.
For GI to work, the density of the protoplanetary disk should excess
the Toomre density \citep{toomre-1964}.
At the same time, it should be low enough to allow efficient cooling, which
is required for a disk to fragment into bound clumps \citep{gammie-2001}.
Both radiative and convective cooling rates may not
be efficient enough for direct formation of giant planets by GI within several tens
of AU from the parent star \citep{rafikov-2005,rafikov-2007}.
As far as the standard core accretion
scenario is concerned, the only possibility to explain the formation of planets with several
Jupiter masses would be to admit that they have formed closer to the star and then were
scattered gravitationally to wider orbits \citep[e.g.][]{veras-armitage-2004}.
Nevertheless, it is difficult to find a mechanism that has circularized their orbits
subsequently. Alternatively, massive planets formed by core accretion could have
smoothly migrated from their birth places outward.
However, this would require
displacing a comparable~-- and  therefore an unrealistically large~-- mass of planetesimals
inward over large distances, rendering this mechanism problematic.
Thus, by and large it remains unclear how the planets have formed.
The era of directly imaged
extrasolar planets that has just begun must eventually bring answers to these
and many other questions.

\begin{acknowledgements}

We thank the referee, Daniel Fabrycky, for an insightful and very useful review,
which greatly helped to improve the paper. AK and RN acknowledge discussions
with Cesary Migaszewski on the dynamics of planets.
SF and SM are funded by the graduate student fellowships of the Thuringia State.
TOBS acknowledges support from {\em Evangelisches Studienwerk e.V. Villigst}.
Part of this work was supported by the
\emph{Deut\-sche For\-schungs\-ge\-mein\-schaft, DFG\/} project numbers
Kr~2164/5-1, Kr~2164/8-1, Ne~515/13-1, Ne~515/13-2, and Ne~515/23-1,
by the {\em Deutscher Akademischer Austauschdienst}
(DAAD), project D/0707543, and by the International Space Science Institute
in Bern, Switzerland (``Exozodiacal Dust Disks and Darwin'' working group,
http://www.issibern.ch/teams/exodust/).

\end{acknowledgements}




\end{document}